\documentclass{aa}

\usepackage[dvipsnames]{xcolor} 
\usepackage{graphicx, subfigure, amsmath, pifont}
\usepackage{amssymb}
\usepackage{multicol}
\usepackage[breaklinks=true]{hyperref} 
\hypersetup{colorlinks=true,citecolor=Blue}
\usepackage{natbib}
\bibpunct{(}{)}{;}{a}{}{,} % to follow the A&A style
\usepackage[english]{babel}
\usepackage{blindtext}
\usepackage[normalem]{ulem}
\usepackage{comment}
\usepackage{lmodern}
\usepackage{ulem} 
\usepackage{soul}

\newcommand{\mtot}{$M_{\text{total}}$}

\graphicspath{{./}{figures/}}
\graphicspath{{./}{figures/FARGO/}}

\begin{document} 

 \title{The morphology of CS\,Cha circumbinary disk suggesting the existence of a Saturn-mass planet}

   \author{
    N.~T.~Kurtovic\inst{\ref{mpia},\ref{das}},
    P.~Pinilla\inst{\ref{mpia},\ref{uclondon}},
    Anna~B.~T.~Penzlin\inst{\ref{tubingen}}, 
    M.~Benisty\inst{\ref{cnrs},\ref{ipag}},
    L.~P\'erez\inst{\ref{das},\ref{npf}}, 
    C.~Ginski\inst{\ref{amsterdam}},
    A.~Isella\inst{\ref{rice}}, 
    W.~Kley\inst{\ref{tubingen}}$^\dag$,
    F.~Menard\inst{\ref{ipag}},
    S.~P\'erez\inst{\ref{usach},\ref{ciras}},
    A.~Bayo\inst{\ref{uvalpo},\ref{npf}}.
   }
   \institute{
   Max-Planck-Institut f\"{u}r Astronomie, K\"{o}nigstuhl 17, 69117, Heidelberg, Germany, \email{kurtovic@mpia.de} \label{mpia}
   \and Departamento de Astronom\'ia, Universidad de Chile, Camino El Observatorio 1515, Las Condes, Santiago, Chile. \label{das}
   \and Mullard Space Science Laboratory, University College London, Holmbury St Mary, Dorking, Surrey RH5 6NT, UK. \label{uclondon}
   \and Institut f\"ur Astronomie und Astrophysik, Universität T\"ubingen, Auf der Morgenstelle 10, D-72076, Germany. \label{tubingen}
   \and Unidad Mixta Internacional Franco-Chilena de Astronom\'{i}a (CNRS UMI 3386), Departamento de Astronom\'{i}a, Universidad de Chile, Camino El Observatorio 33, Las Condes, Santiago, Chile \label{cnrs}
   \and Univ. Grenoble Alpes, CNRS, IPAG, F-38000 Grenoble, France. \label{ipag}
   \and N\'ucleo Milenio de Formaci\'on Planetaria (NPF), Chile. \label{npf}
   \and Anton Pannekoek Institute for Astronomy, University of Amsterdam, Science Park 904,1098XH Amsterdam, The Netherlands. \label{amsterdam}
   \and Department of Physics and Astronomy, Rice University, 6100 Main Street, MS-108, Houston, TX 77005, USA. \label{rice}
   \and Departamento de F\'isica, Universidad de Santiago de Chile, Av. Victor Jara 3659, Santiago. \label{usach} 
   \and Center for Interdisciplinary Research in Astrophysics and Space Exploration (CIRAS), Universidad de Santiago de Chile, Estaci\'on Central, Chile. \label{ciras}
   \and Instituto de F\'isica y Astronom\'ia, Facultad de Ciencias, Universidad de Valpara\'iso, Av. Gran Breta\~na 1111, Playa Ancha, Valpara\'iso, Chile. \label{uvalpo}
   }
   \date{}

% F.M. 0000-0002-1637-7393

 \authorrunning{N.~Kurtovic}
 \titlerunning{CS\,Cha Circumbinary disk}
% \abstract{}{}{}{}{} 
% 5 {} token are mandatory

  \abstract
  % context heading (optional)
   {Planets have been detected in circumbinary orbits in several different systems, despite the additional challenges faced during their formation in such an environment.}
  % aims heading (mandatory)
   {We investigate the possibility of planetary formation in the spectroscopic binary CS\,Cha by analyzing its circumbinary disk.}
  %methods heading (mandatory)
   {The system was studied with high angular resolution ALMA observations at 0.87\,mm. Visibilities modeling and Keplerian fitting are used to constrain the physical properties of CS\,Cha, and the observations were compared to hydrodynamic simulations.}
  % results heading (mandatory)
   {Our observations are able to resolve the disk cavity in the dust continuum emission and the $^{12}$CO J:3-2 transition. We find the dust continuum disk to be azimuthally axisymmetric (less than $9\%$ of intensity variation along the ring) and of low eccentricity (of 0.039 at the peak brightness of the ring). }
  % conclusions heading (optional), leave it empty if necessary  {}
  {Under certain conditions, low eccentricities can be achieved in simulated disks without the need of a planet, however, the combination of low eccentricity and axisymmetry is consistent with the presence of a Saturn-like planet orbiting near the edge of the cavity.}


  \maketitle

\section{Introduction}                  \label{sect:intro}

Over the last decade, space telescopes such as Kepler and the Transiting Exoplanet Survey Satellite (TESS) have successfully detected several planets in circumbinary orbits, which are also known as P-type orbit planets \citep[see ][]{doyle2011, kostov2020}. These planets have been found to share some orbital properties, such as: i) most of them are located close to the inner dynamical stability limit \citep{dvorak1986, holman1999, martin2019} and ii) their orbits are mostly coplanar and of low eccentricity, with a planet occurrence rate similar to single stellar systems \citep[][]{armstrong2014, martin2014}. These common characteristics cannot be explained as simply observational biases \citep{martin2014}, which could be evidence that common formation mechanisms are at play for these planets.

Due to the interaction between the two central stars, not all the regions of a circumbinary disk are suitable for planet formation. Tidal forces are expected to carve a central cavity in the disks, where the material density is severely reduced \citep{artymowicz1994, miranda2015}, and oscillations in the eccentricity of the orbits make extremely challenging to have planetesimal and pebble accretion in the regions close or within the dynamical stability limit \citep{paardekooper2012, pierens2020}. 
Consequently, the detection of several planets in the edge of that region suggests that the planets were formed farther away and later migrated to the location where they can now be observed \citep{pierens2007, meschiari2012, kley2014, thun2018}.

Hydrodynamic simulations of circumbinary disks have shown that disks become eccentric due to dynamical instabilities, and the properties of the cavity will be dependent on the binaries and disk itself \citep[see][]{lubow1991, macfadyen2008, thun2017, hirsh2020, munoz2020, ragusa2020}. 
The inclusion of a planet can disrupt this behavior, as the gap opened by a planet can shield the outer disk from the action of the binaries, allowing it to become more circular \citep{kley2019, penzlin2021}. Therefore, the study of a disk kinematics and structures of a young circumbinary disk could either hint at or exclude the presence of such planets.

A particularly interesting multiple stellar system is CS\,Cha, a spectroscopic binary with a period of at least 7\,yr \citep{guenther2007,nguyen2012} and a member of Chameleon I association, with an estimated age of $4.5\pm1.5\,$Myr \citep{luhman2007}. CS\,Cha is located at 169\,pc estimated from the parallax of GAIA EDR3 \citep{gaia2016b, gaia2021edr3} and the combined luminosity of the binary is estimated to be $L_\star=1.45\,L_\odot$ \citep{manara2014}. 
The system is known to host a circumbinary disk, which was first identified from its spectral energy distribution (SED) due to an excess in the infrared wavelengths \citep{gauvin1992}, and later detected at 1.3\,mm wavelength \citep{henning1993}. The system was cataloged as a transition disk due to its SED shape, which was modeled early on as a disk with a central cavity \citep{espaillat2007}. Over the last two decades, there have been several attempts to measure the cavity size and ring location, mainly through its SED \citep{espaillat2007, kim2009, espaillat2011, ribas2016}, with the latest estimations being $R_{\rm{cav}}=18_{-5}^{+6}$\,au.
Recent observations with the Spectro-Polarimetric High-contrast Exoplanet REsearch (\emph{SPHERE}) at the Very Large Telescope (\emph{VLT}) made it possible to spatially resolve the disk scattered light, demonostrating that if there is a cavity in scattered light emission (small micron-sized grains), it must be within the coronagraph hidden region, setting an upper limit of 15.6\,au \citep{ginski2018}. 
Finally, the modeling of interferometric data of the millimeter dust continuum emission with a 1D radial profile, suggests that the disk has a ring-like shape with its peak located at $204\pm7$\,mas ($34.5\pm1.2\,$au) \citep{norfolk2021}.

Combined observations of NAOS-CONICA (\emph{NACO}) at the VLT, SPHERE, and the Hubble Space Telescope (\emph{HST}), have allowed the identification of a co-moving companion located at $\approx1.3''$ ($\approx220\,$au) of projected distance to the CS\,Cha binaries \citep{ginski2018}. Initially, it was thought to be a planetary mass object \citep{ginski2018}, however, its optical and near-infrared (NIR) spectra have shown that it is possible that CS\,Cha\,B is actually an M-dwarf star severely obscured by a highly inclined disk and outflows \citep{haffert2020}.
Such a circumstellar environment on CS\,Cha\,B is also supported by a very high degree of polarization observed with SPHERE and by the detection of a mass accretion rate of $\dot{M}=4\cdot10^{-11\pm0.4}\,M_\odot\,yr^{-1}$ \citep{haffert2020}. 

Motivated by the detection of CS\,Cha\,B, the system was observed by Atacama Large Millimeter Array (ALMA) with the aim of characterizing this newly detected companion. These observations also provide one of the deepest and highest sensitivity observations available for a Class II circumbinary disk.
In the present work, we analyze the high angular resolution millimeter observations of the CS\,Cha system, which contains the dust continuum emission at 0.87\,mm and $^{12}$CO J:3-2 molecular line emission. The observation details and calibration applied to the data are described in Section~\ref{sec:observations}. The analysis of the observations and their modeling is presented in Section~\ref{sec:results}, while an attempt to constrain the physical mechanisms responsible of the observed emission structures is detailed in Section~\ref{sec:hydro_sim}. We discuss our results in Section~\ref{sec:discussion}, and then we summarize the main conclusions of our work in Section~\ref{sec:conclusions}.

\section{Observations} \label{sec:observations}

% Data description
This work includes 0.87\,mm observations of the circumbinary system CS\,Cha, observed with ALMA Band 7 as part of the ALMA project 2017.1.00969.S (PI: M.~Benisty) between 26-Nov-2017 and 12-Dec-2017. The correlator was configured to observe four spectral windows: three covered dust continuum emission centered at $334.772\,$GHz, $336.600\,$GHz, and $347.471\,$GHz, with a total bandwidth of $2\,$GHz; the remaining one was centered at $345.770\,$GHz to observe the molecular line $^{12}$CO in the J:3-2 transition (from now on referred to as $^{12}$CO) with a frequency resolution of $122.07$\,kHz ($\sim 0.1\,$km\,s$^{-1}$ per channel). The total time on source was 273.1\,min, spanning baselines from 15.1\,m to 8547.6\,m from ALMA antenna configurations C43-8 and C43-7.

% Continuum self calibration
We started from the pipeline calibrated data, after executing the \texttt{scriptforPI} provided by ALMA. Then, using \texttt{CASA 5.6.2}, we extracted the dust continuum emission from the spectral window targeting $^{12}$CO, by flagging the channels located at $\pm 25\,$km\,s$^{-1}$. The remaining channels were combined with the other continuum spectral windows to obtain a ``pseudo-continuum'' dataset, and we averaged them into 125\,MHz channels and $6\,$s bins to reduce data volume. To enhance the signal to noise ratio (S/N), self-calibration was applied on the continuum. We used a Briggs robust parameter of 0.5 for the imaging of the self-calibration process, and we applied four phase and one amplitude calibrations, using the whole integration time as the solution interval for the amplitude calibration and first phase calibration, while for the remaining phase calibrations, we used 360\,s, 150\,s, and 60\,s. 

% Imaging + JvM
After self-calibration was complete, we explored different alternative values for the robust parameter to image the data using the \texttt{CLEAN} algorithm.
For the \texttt{multiscale} parameter, we used  (0$\times$, 0.2$\times$, 0.5$\times$, 1$\times$) the beam size, which, in combination with a \texttt{smallscalebias} of 0.45, returns a smoother model for the emission. This value for \texttt{smallscalebias} is smaller than the default 0.6, which leads to the algorithm preferring the extended scales before point sources. To avoid introducing PSF artifacts that could be mistaken for faint emission, we lowered the \texttt{gain} parameter to $0.05$ (it controls the fraction of the flux that is cleaned in every iteration), and increased the \texttt{cyclefactor} to $1.5$ (it controls the frequency with which major clean cycles are triggered), both of them chosen for a more conservative imaging compared to the default values\footnote{Check \url{https://casa.nrao.edu/docs/taskref/tclean-task.html} for a description of the parameters}. We cleaned down to a 4$\sigma$ threshold, and applied the \emph{JvM correction} to our images, which accounts for the volume ratio $\epsilon$ between the point spread function (PSF) of the images and the restored Gaussian of the \small{\texttt{CLEAN}} beam, as described in \citet{jorsater1995} and \citet{czekala2021}.

% Gas calibration
The calibration tables obtained from the dust continuum self-calibration were applied to the molecular line emission channels, and then the continuum emission was subtracted from them with the \texttt{uvcontsub} task. To increase the S/N of the images, we imaged the $^{12}$CO channels with a lowered velocity resolution of 0.25\,km\,s$^{-1}$, centered at 3.65\,km\,s$^{-1}$, which is approximately the velocity at the local standard of rest (VLSR). Different robust parameters ranging from -1.2 to 1.2 were explored to find the best trade-off between angular resolution and sensitivity. Additionally, we also applied uv tapering\footnote{The term ``uv'' in this work is used to refer to the visibility plane.} to generate another set of $^{12}$CO images with a more circularized beam.

The JvM correction was also applied to the CO channel maps before any analysis was carried out on them. The package \texttt{bettermoments} \citep{bettermoments, teague2019} was used to create additional image products from the channel maps. This package fits a quadratic function to find the peak intensity of the line emission in each pixel, and the velocity associated with it, but we also used it to generate the moment 0 and moment 1 of each velocity cube. All the moment images were clipped at 3 sigma and no mask was used.

% data treatment for uvmodeling
To accurately analyze the observations, we also applied uv modeling to the continuum visibilities of the source, as described in Section~\ref{sec:cont_morph}. To further reduce the data volume after finishing the self-calibration, we averaged the continuum emission into 1 channel per spectral window \citep[as in][]{andrews2021} and 30\,s of time binning. We used each binned channel central frequency to convert the visibility coordinates into wavelength units.

%%%%%%%%%%%%%%%%%%%%%%%%%%%%%%%%%%%%%%%%%%%%%%%%%%%%%%%%%%%%%%%%%%%%%%%%%%%%%%%%

%% Data Analysis.
\section{Observational results} \label{sec:results}

\subsection{Circumbinary disk: Dust continuum emission physical properties}

\begin{figure*}[t]
 \centering
        \includegraphics[width=18cm]{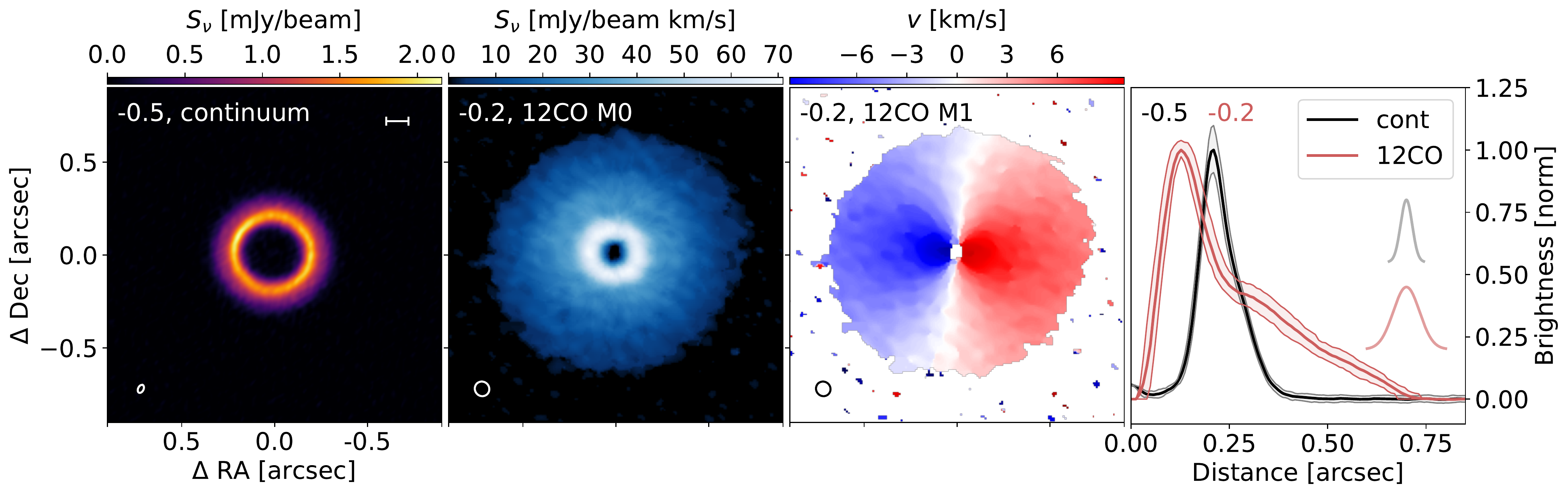}\\ \vspace{-0.1cm}
   \caption{Reconstructed images of dust continuum emission and $^{12}$CO. From left to right: Dust continuum emission from CS\,Cha imaged with a robust parameter of -0.5, moment 0 and moment 1 of the $^{12}$CO imaged with a robust of -0.2, and radial profiles for the continuum and $^{12}$CO emission calculated by deprojecting the images with the inclination and position angle of Model 2e (see Section~\ref{sec:cont_morph}). The ellipse in the left bottom corner of the panels represents the synthesized beam of the images, which is $30\times46\,$mas for the dust continuum and $80\times77\,$mas for the $^{12}$CO. The scale bar in the top right of the first panel represents 20\,au at the distance of the source. The Gaussians in the right panel represent the average radial resolution of the profiles, and in the same panel, the colored region in the profiles represent the $1\,\sigma$ dispersion at each radial location.}
   \label{fig:obs_images}
\end{figure*}

The millimeter emission is resolved into a single disk around the binary stars (a circumbinary disk), as shown in Fig.~\ref{fig:obs_images} after being imaged with a robust parameter of -0.5, returning an angular resolution of $30\times46$\,mas. At a nominal resolution ($61\times87$\,mas with a robust parameter of 0.5), the disk appears as a single smooth ring with a central cavity, however, higher angular resolution images resolve the disk radial structure, showing evidence of a radially asymmetric ring (right panel of Fig.~\ref{fig:obs_images}). A gallery with the dust continuum emission reconstructed with different robust parameters ranging from -1 to 1 is included in Fig.~\ref{fig:continuum_gallery} in the appendix. For continuum images with robust parameters larger than 1, the sensitivity changes are negligible, as the beam size increases to an extent less than 10\% and the point spread function is poorer due to the sparser uv coverage at short baselines compared to long baselines, resulting in stronger sidelobes and, thus, stronger structured residuals.

The radial profiles were obtained by deprojecting the images with the geometry parameters obtained in Sect.~\ref{sec:cont_morph} (inc=17.86\,deg and PA=82.6\,deg, see Tab.~\ref{tab:mcmc_results}), where we considered multiple Gaussian components and eccentricities to describe the circumbinary disk.
We find that the dust continuum ring profile peaks at $205\pm5\,$mas from the disk center, which is $34.6\pm0.8\,$au at the distance of the source. Since it was calculated from the image, we used 5\,mas as a conservative uncertainty (the pixel size), which is consistent with the previous study by \citet{norfolk2021}.

In order to estimate the optical depth $\tau$ of the emission, we followed the same approach as in \citet{pinilla2021}, assuming that the disk emits as a black body and therefore $\tau=-\ln(1 - T_{B}/T_{\text{phys}})$, where $T_{B}$ is the brightness temperature, and $T_{\text{phys}}$ is the physical temperature of the midplane. We estimated the $T_{B}$ from the different dust continuum images by starting from the Rayleigh-Jeans approximation. When the beam size is increased (by using larger robust parameters), the emission becomes more diluted and, so, the peak temperatures decreases. For the image with a robust parameter of 0.5, the peak brightness temperature of the image reaches $12.3\pm0.1$\,K (brightness temperature uncertainty given with 3\,sigma confidence), while for the image with robust value of -1.0, it reaches $17.4\pm0.4$\,K, since the ring is better resolved. For this reason, we decided to use the image generated with robust -0.5 to estimate the optical depth, given it has a high S/N and also high spatial resolution. From this image, we obtained a peak $T_{B}$ of $16.1\pm0.3$\,K.

For the $T_{\text{phys}}$, we need additional assumptions. If we consider the midplane temperature to be at the standard 20\,K, then we find a peak optical depth of $\tau_{\text{peak}}=1.34$. On the other hand, if we consider the approximated luminosity-dependent temperature relation from \citet{andrews2013}, $T=25(L_\star/L_\odot)^{0.25}\,$K, and $L_\star=1.45\,L_\odot$ for the stellar luminosity \citep{manara2014}, then we can estimate $T_{\text{phys}}=27.4$\,K and $\tau_{\text{peak}}=0.77$. Both estimates should be considered with caution, as the first assumes a single constant temperature and the latter comes from a luminosity relation for disks with a single stellar host.

We calculated the dust mass of the model by assuming that the flux ($F_\nu$) received has a wavelength of 0.87\,mm and is being emitted by optically thin dust with a constant temperature of 20\,K \citep[as in ][]{ansdell2016, pinilla2018}, and, alternatively, with a constant temperature of $27.4$\,K. In both approaches we follow \citet{hildebrand1983}:
\begin{equation}
    M_{\text{dust}} = \frac{d^2\,F_\nu}{\kappa_\nu\,B_\nu(T(r))} \text{,}
\label{eq:mdust}
\end{equation}
\noindent where $d$ is the distance to the source, $\nu$ is the observed frequency, $B_\nu$ is the Planck function at the frequency $\nu$, and $\kappa_\nu=2.3(\nu/230\,\text{GHz})^{0.4}\,\text{cm}^{2}\text{g}^{-1}$ is the frequency-dependent mass absorption coefficient \citep[as in][]{andrews2013}. The total flux from the source is estimated by taking the weighted average of the baselines shorter than 28\,k$\lambda$, which gives $F_\nu=180.2\pm0.5$\,mJy, not accounting for the $10\%$ uncertainty of ALMA fluxes. We chose to measure it from the visibilities that do not resolve the disk emission to avoid introducing additional uncertainties related to image reconstruction and possible dependence on the mask chosen. Replacing this value in Eq.~\ref{eq:mdust}, we obtain a dust mass of $69.0\pm0.1\,M_\oplus$ when assuming $T_{phys}=20\,$K, and $44.7\pm0.1\,M_\oplus$ for $T_{phys}=27.4\,$K.% 
Therefore, the dust mass content is uncertain either because of the temperature assumption and the poor constraints that we have on the dust opacities from the observations.

\subsection{No detection of emission near CS\,Cha\,B}

\begin{figure}[t]
 \centering
        \includegraphics[width=8.5cm]{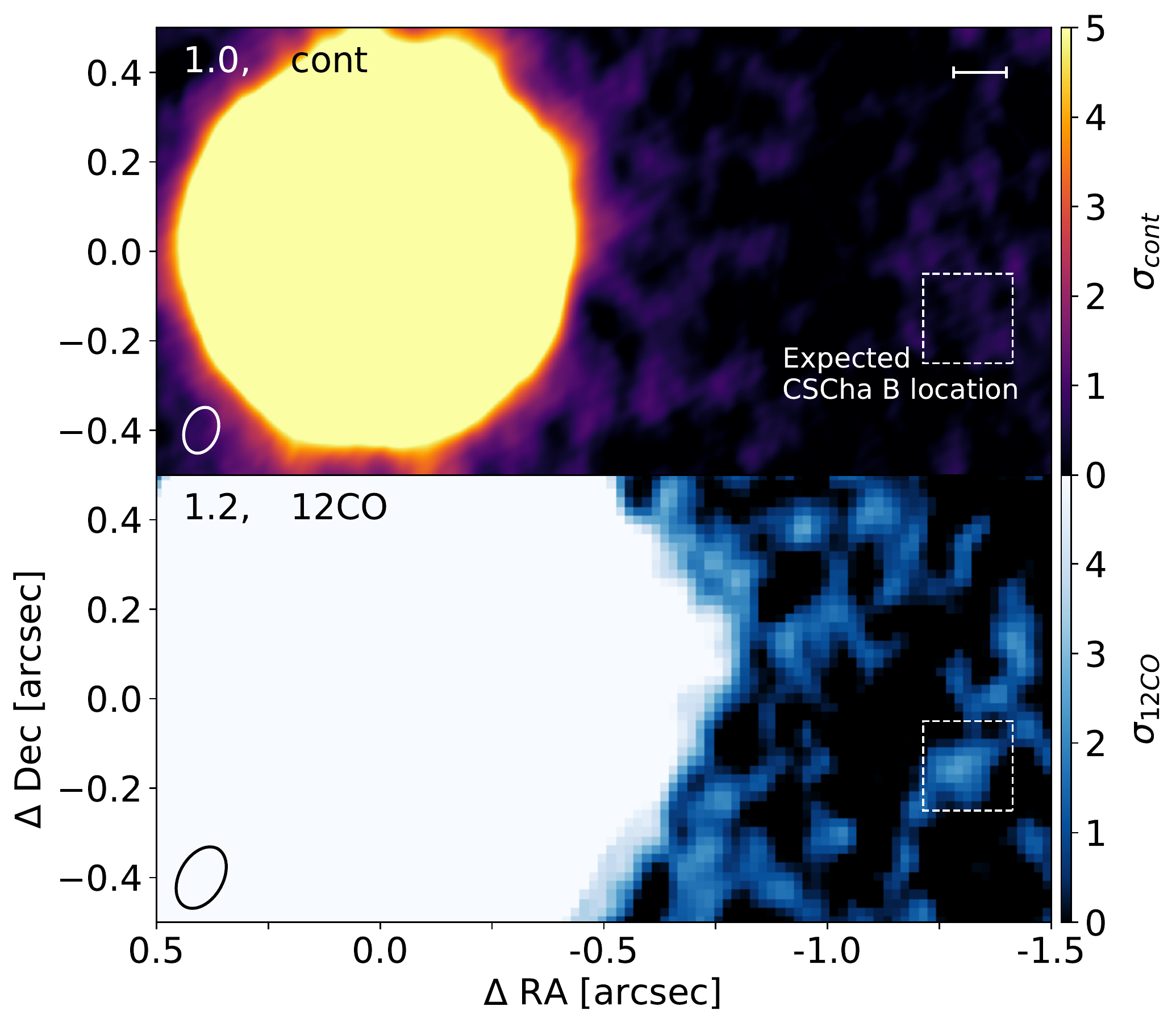}\\ 
        \vspace{-0.1cm}
   \caption{High-sensitivity millimeter emission images of CS\,Cha. \textbf{Upper panel}: Dust continuum image generated with a robust parameter of 1.0. The color scale is linear and has been saturated to show the emission between $0$ and $5\sigma_{\text{cont}}$, with $\sigma_{\text{cont}}=11.8\,\mu$Jy/beam being the rms of this image. A box of $0.2''$ per side is centered at the expected location of CS\,Cha\,B. The beam size is $105\times75\,$mas, and is shown in the lower left corner of the figure. The scale bar at the top right represents 20\,au. \textbf{Lower panel}: $^{12}$CO emission image generated with a robust parameter of 1.2, after stacking all the channel maps between -24 and 24\,km\,s$^{-1}$ around the rest frame. The beam size is $148\times97\,$mas, and is shown in the lower left corner of the figure. The color scale is linear and has been saturated to show the emission between $0$ and $5\,\sigma_{\text{12CO}}$, with $\sigma_{\text{12CO}}=1.4\,$mJy/beam being the rms of this image.}
   \label{fig:cschab}
\end{figure}

We did not detect any significant emitting source at the expected location of CS\,Cha\,B, neither in dust continuum emission nor $^{12}$CO, as shown in the upper and lower panels of Fig.~\ref{fig:cschab}, respectively, where the emission has been saturated to $5\sigma$ of each image. In the dust continuum, by using our highest sensitivity image (generated with a robust parameter of 1.0) and based on the assumption that CS\,Cha\,B is a point source, we can estimate a 3$\sigma$ upper limit for millimeter emission to be $35.4\mu$Jy. This emission translates into a dust mass upper limit of $M_B<0.015\,M_\oplus$ under the assumption of 20\,K and optically thin emission. 
Even if the disk is not a compact source, the beam size of the robust 1.0 image is $\approx18\times13\,$au at the distance of the source, therefore, the dust disk would have been unresolved even if it had a size of $10\,$au. 

In $^{12}$CO, we do not detect any significant emission at the location of CS\,Cha\,B either and this non-detection is independent from the channel map velocity width and synthesized beam size used for image reconstruction. As a final test for the detection of CS\,Cha\,B, we generated a cube with a robust parameter of 1.2, no uv tapering, and a channel width of 1\,km\,s$^{-1}$, going from -24 to 24\,km\,s$^{-1}$ around the rest frame of the $^{12}$CO line. These channels were all stacked and the result is displayed in the lower panel of Fig.~\ref{fig:cschab}. The peak emission within the square mask does not reach a significance of $2\sigma$.

\subsection{Dust morphology from the visibility fitting}\label{sec:cont_morph}

To precisely constrain the structure of the dust continuum emission, we applied uv modeling to the visibilities of the source, via parametric models. In principle, the brightness profile ($f$) of an axisymmetric disk would only depended on the radial distance to the center of the disk (given by $f:=f(r)$). However, circumbinary disks are expected to display some eccentricity due to the interaction between the disk material and the binaries \citep[e.g.,][]{thun2017,kley2019}, and so, it is convenient to define the brightness profile not as a function of the radius, but as a function of the semi-major axis instead ($f:=f(a)$). We calculated the eccentric coordinate system by following the same approach that \citet{marino2019} and \citet{booth2021}:
\begin{equation}
    a(r, \phi) \, = \, r \, \frac{1\,-\,e\cos(\phi-\omega)}{1\,-\,e^2} \text{,}
\end{equation}
\noindent where the semi-major axis $a$ is a function of the radial distance from the center of mass and the azimuthal angle ($r, \phi$), and it can also be modified by the eccentricity, $e,$ and the argument of the periastron $\omega$. It is pertinent to notice that this coordinate system does allow for the solution $e=0.0$, which returns the standard polar coordinates.

Several models with increasing complexity have been considered to describe the disk around CS\,Cha, which are composed of a combination of Gaussians shapes by $f=\sum_i g_i$, where $g_i$ is the $i$th Gaussian. Each subsequent model is motivated by the residuals of the best previous model, but they all share the same basic shape for the disk, described by a bright Gaussian ring ($g_0$) for the inner side of the ring emission, plus a radially asymmetric Gaussian ring ($g_1$) to describe the outer side of the ring emission. This $g_1$ component has a different width for each side of its peak, also known as broken-Gaussian ($(\sigma_i, \sigma_o)$ for the inner and outer part, respectively). The additional features considered in the more complex models were a centrally peaked Gaussian ($g_2$) and an extended Gaussian ring ($g_3$). All these components are schematized in Fig.~\ref{fig:schematic_profiles} in the appendix, and the ones considered in each model are: 
\begin{enumerate}
    \item Model 2g, composed of $g_0$+$g_1$ with eccentricity and argument of the periastron of ($e_0, \omega_0$); 
    \item Model 3g, composed of $g_0$+$g_1$+$g_2$ with ($e_0, \omega_0$); 
    \item Model 4g, composed of $g_0$+$g_1$+$g_2$+$g_3$ with ($e_0, \omega_0$);
    \item Model 2e, composed of $g_0$+$g_2$ with ($e_0, \omega_0$), and $g_1$+$g_3$ with ($e_1, \omega_1$).
\end{enumerate}

\begin{table*}[t]
\centering
\begin{tabular}{ c|c|c|c|c||c||c } 
  \hline
  \hline
\noalign{\smallskip}
    Component   &  & Model 2g    & Model 3g    & Model 4g    & \textbf{Model 2e}    & Units \\
\noalign{\smallskip}
  \hline
  \hline
\noalign{\smallskip}
geometry& $\delta_{\rm{RA}}$  & $-13.17_{-0.06}^{+0.04}$ & $-13.28_{-0.01}^{+0.09}$ & $-13.18_{-0.04}^{+0.06}$ & $-13.16_{-0.03}^{+0.08}$ & mas \\
        & $\delta_{\rm{Dec}}$ & $ 2.28_{-0.08}^{+0.03}$  & $ 2.07_{-0.04}^{+0.01}$  & $ 2.35_{-0.05}^{+0.06}$  & $ 1.37_{-0.03}^{+0.08}$  & mas \\
        & inc                 & $17.78_{-0.05}^{+0.01}$  & $17.95_{-0.04}^{+0.01}$  & $17.79_{-0.03}^{+0.02}$  & $17.86_{-0.01}^{+0.05}$  & deg \\
        & PA                  & $82.6$ fixed             & $82.6$ fixed             & $82.6$ fixed             & $82.6$ fixed             & deg \\
\noalign{\smallskip}
  \hline
\noalign{\smallskip}
eccentricity     & $e_0$       & $0.023_{-0.001}^{+0.001}$ & $0.024_{-0.001}^{+0.001}$ & $0.023_{-0.001}^{+0.001}$ & $0.039_{-0.001}^{+0.001}$ & - \\
     & $\omega_0$    & $-5.11_{-0.55}^{+0.35}$   & $-5.65_{-0.12}^{+0.78}$   & $-5.48_{-0.43}^{+0.46}$   & $-1.02_{-0.28}^{+0.47}$   & deg \\
\noalign{\smallskip}
     & $e_1$       &                           &                           &                           & $0.019_{-0.001}^{+0.001}$ & - \\
     & $\omega_1$    &                           &                           &                           & $-8.42_{-0.27}^{+0.91}$   & deg \\
  \hline
\noalign{\smallskip}
$g_0$& $f_0$         & $23.05_{-0.16}^{+0.02}$   & $24.99_{-0.03}^{+0.14}$   & $25.15_{-0.11}^{+0.05}$   & $24.61_{-0.12}^{+0.09}$   & ($\mu$Jy/pix) \\
     & $r_0$         & $203.3_{-0.1}^{+0.1}$     & $203.0_{-0.1}^{+0.1}$     & $203.7_{-0.1}^{+0.1}$     & $202.7_{-0.1}^{+0.1}$     & mas \\
     & $\sigma_{0}$  & $16.6_{-0.3}^{+0.1}$      & $20.3_{-0.1}^{+0.3}$      & $20.0_{-0.1}^{+0.1}$      & $18.8_{-0.1}^{+0.2}$      & mas \\
  \hline
\noalign{\smallskip}
$g_1$&$f_1$          & $16.72_{-0.04}^{+0.20}$   & $14.46_{-0.19}^{+0.02}$   & $14.31_{-0.07}^{+0.11}$   & $15.21_{-0.14}^{+0.10}$   & ($\mu$Jy/pix) \\
    & $r_1$          & $238.7_{-1.3}^{+0.3}$     & $255.4_{-0.1}^{+0.1}$     & $257.4_{-0.8}^{+0.5}$     & $251.1_{-0.7}^{+0.9}$     & mas \\
    & $\sigma_{1i}$  & $56.8_{-0.8}^{+0.2}$      & $58.6_{-0.1}^{+0.7}$      & $63.0_{-0.5}^{+0.4}$      & $58.8_{-0.4}^{+0.5}$      & mas \\
    & $\sigma_{1o}$  & $55.9_{-0.1}^{+0.4}$      & $46.3_{-0.5}^{+0.1}$      & $44.9_{-0.3}^{+0.4}$      & $47.8_{-0.4}^{+0.4}$      & mas \\
  \hline
\noalign{\smallskip}
$g_2$& $f_2$         &                           & $0.59_{-0.01}^{+0.01}$    & $2.81_{-0.31}^{+2.18}$    & $6.10_{-3.33}^{+0.03}$    & ($\mu$Jy/pix) \\
     & $\sigma_{2}$  &                           & $262.4_{-1.2}^{+2.2}$     & $12.3_{-4.7}^{+1.8}$      & $10.2_{-0.4}^{+5.5}$      & mas \\
  \hline
\noalign{\smallskip}
$g_3$& $f_3$         &                           &                           & $0.37_{-0.02}^{+0.02}$    & $0.37_{-0.02}^{+0.02}$    & ($\mu$Jy/pix) \\
     & $r_3$         &                           &                           & $214.2_{-11.8}^{+14.9}$   & $161.5_{-11.1}^{+16.7}$   & mas \\
     & $\sigma_{3}$  &                           &                           & $179.2_{-5.9}^{+5.2}$     & $206.3_{-8.1}^{+5.4}$     & mas \\
\noalign{\smallskip}
  \hline
  \hline
\noalign{\smallskip}
     & $F_\lambda$   & $178.82\pm0.02$           & $180.70\pm0.04$           & $180.49\pm0.04$           & $180.57\pm0.05$           & mJy \\
\noalign{\smallskip}
  \hline
  \hline
\end{tabular}
\caption{Best parameters from the uv modeling; ``mas'' stands for milliarcsecond.}
\label{tab:mcmc_results}
\end{table*}

\begin{figure*}[t]
 \centering
        \includegraphics[width=18cm]{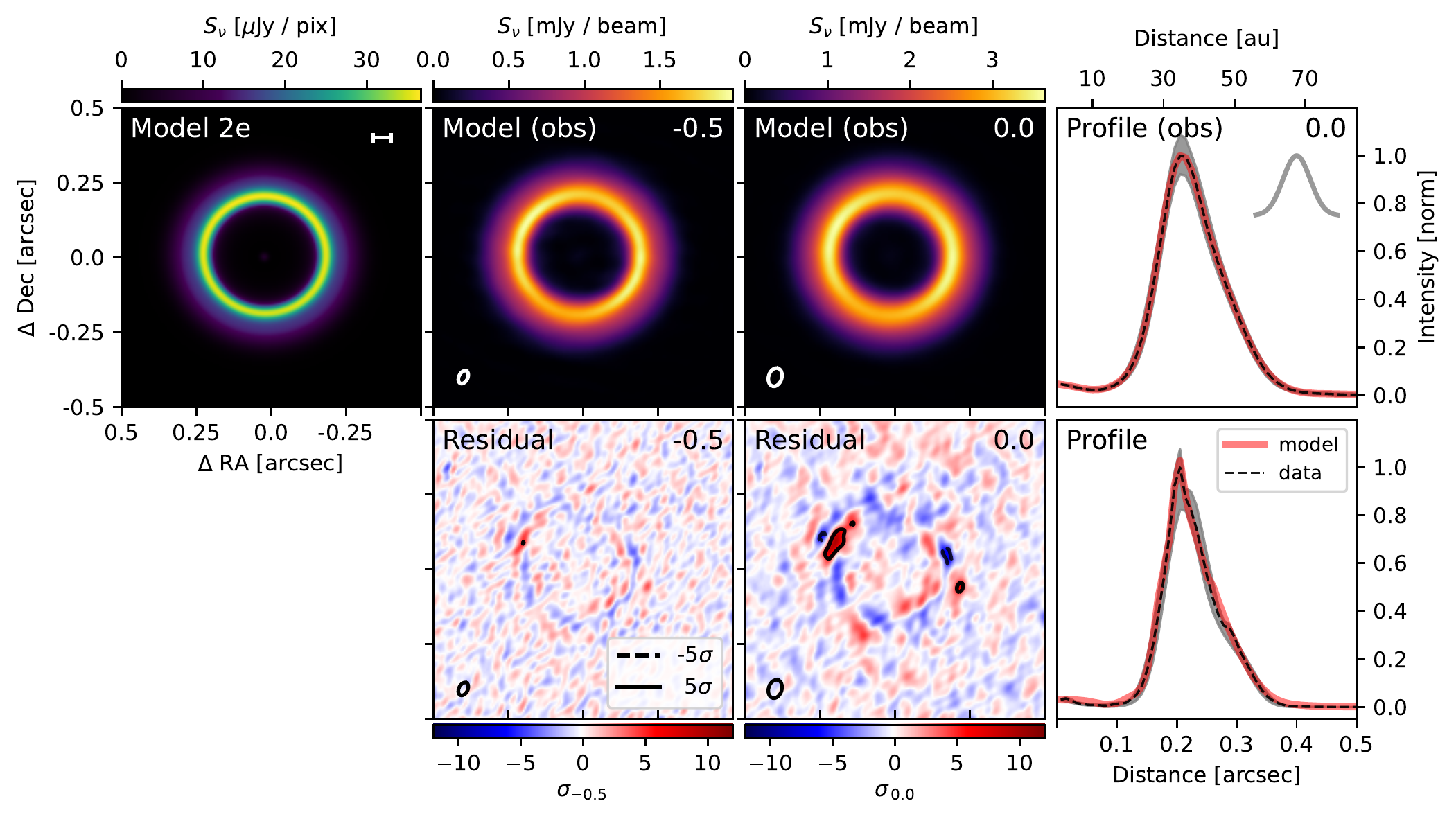}\\ \vspace{-0.1cm}
   \caption{Best solution for the dust continuum emission generated using the Model 2e, which considers two eccentricities for the disk components. \textbf{Upper row}: Left panel shows the synthetic image of the best model found. Middle panels show how this model would have been observed by ALMA with two different robust parameters, comparable to the images from Fig.~\ref{fig:continuum_gallery}. Right panel shows the radial profile obtained from the beam convolved images generated with a robust parameter of 0.0, and the average beam resolution shown with a Gaussian in gray. \textbf{Lower row}: Middle panels show the residuals left by the best model, imaged with two different robust parameters shown in the upper left corner. Right panel shows the intensity profile of the model obtained from \texttt{tclean} and the best Model 2e (not convolved by beam).}
   \label{fig:mod_two_ecc}
\end{figure*}

The CS\,Cha disk is close to being face-on (as seen in Fig.~\ref{fig:obs_images}) and so, the dust continuum modeling has a strong dependence between the center of the disk ($x_0, y_0$), the inclination (inc), the position angle (PA), and the eccentricity parameters ($e$, $\omega$). To reduce the number of free parameters, we used the $^{12}$CO observations (which are independent from the dust continuum observations) to constrain the PA of the disk. As explained in Section~\ref{sec:co_obs}, the preliminary kinematic fittings to the $^{12}$CO show that it has a position angle of $82.6$\,deg, and so we fix this value in our uv modeling. 

To find the best set of parameters for each different model, we used the package \texttt{emcee} to sample the parameter space with a Markov chain Monte Carlo (MCMC) Ensemble Sampler \citep{emcee2013}, using 250 walkers and a flat prior for all the parameters. We used \texttt{galario} to compute the visibilities of the synthetic images, which were generated with a pixel size of $5$\,mas, as the images from \texttt{tclean}. The total flux of the model is calculated by averaging the real part of the ten shortest baselines (u-v pairs) after convergence, and we picked 5000 MCMC random walkers positions to calculate the uncertainty of the flux.

The best parameters for each model and their uncertainties are summarized in Table~\ref{tab:mcmc_results}. All the models show consistent results for the disk flux ($F_\lambda$), the location of the radius that includes the 68\% and 90\% of the flux ($R_{68}$ and $R_{90}$), and the location of the peak of the ring (given by the parameter $r_0$). The difference between the models can be better observed when the residuals are imaged, as seen in Fig.~\ref{fig:mod_two_ecc} for the Model 2e, and in the Fig.~\ref{fig:model_others} for the Models 2g, 3g, and 4g, in the appendix. The simplest model, Model 2g, shows strong structured residuals in the ring region, and also in the cavity, evidence that the cavity is not completely depleted of dust continuum emission. Then, Model 3g takes into account this inner cavity emission with a Gaussian that peaks at the center of the disk ($g_2$), however, this Gaussian is spread over the whole disk in the attempt to account for an extended diffuse emission, rather than only fitting the cavity emission. To fix this behavior, Model 4g includes a new diffuse extended Gaussian for the ring ($g_3$) in addition to the Gaussian for the inner cavity emission ($g_2$). It ultimately succeeds at describing the cavity emission, but still leaves structured residuals in the circumbinary ring. 

The residuals from the Model 4g subtracted too much flux from some regions (seen in blue in the residual image), and not enough from others (seen in red in the residual image). The structure of these residuals cannot be explained by any combination of offsets in center ($\delta_{\text{RA}}$, $\delta_{\text{Dec}}$), nor geometry (inc, PA), which is discussed in depth in the Appendix of \citet{andrews2021}. To account for the residuals of Model 4g, two additional models were considered: i) a model where the innermost emission has a different inclination compared to the outer most regions ($g_0$+$g_2$ have a inclination inc$_0$, while $g_1$+$g_3$ have another inc$_1$), but they share the same eccentricity; and  ii) a model where those components have the same inclination, but different eccentricities. The model with different inclinations returned residuals that were similar to the ones from Model 4g, and the inclinations (inc$_0$, inc$_1$) were not disparate from the noise level. On the other hand, the model with two eccentricities (Model 2e) was successful in accounting for the structured residuals observed in the circumbinary ring, as seen in Fig.~\ref{fig:mod_two_ecc}.

The best model, Model 2e, found two different eccentricities for the inner most emission and the outermost emission from the circumbinary disk, with the inner ring eccentricity ($e_0=0.39$) being about twice that of the outer disk eccentricity ($e_1=0.19$), as shown in Table~\ref{tab:mcmc_results}. We calculated the mass of the dust in the model by following the same assumptions we took in the dust continuum images: optically thin emission and 0.87\,mm, with dust at 20\,K. The flux from all components of Model 2e adds up to $180.57\pm0.05$\,mJy, which translates into a dust mass of $69.05\pm0.02\,M_\oplus$ at the distance of this source (not considering the $10\%$ uncertainty of ALMA fluxes). As for central Gaussian, $g_2$, alone in the Model 2e, we find a flux of $150\pm16\,\mu$Jy, or a dust mass of $0.057\pm0.006\,M_\oplus$ being detected inside the cavity. This dust mass is $\approx0.5\,M_{\text{Mars}}$, or $\approx5\,M_{\text{Moon}}$, for reference.

\subsection{$^{12}$CO J:3-2 emission}\label{sec:co_obs}

\subsubsection{Emission profile of the $^{12}$CO emission}

\begin{figure*}[t]
 \centering
        \includegraphics[width=15cm]{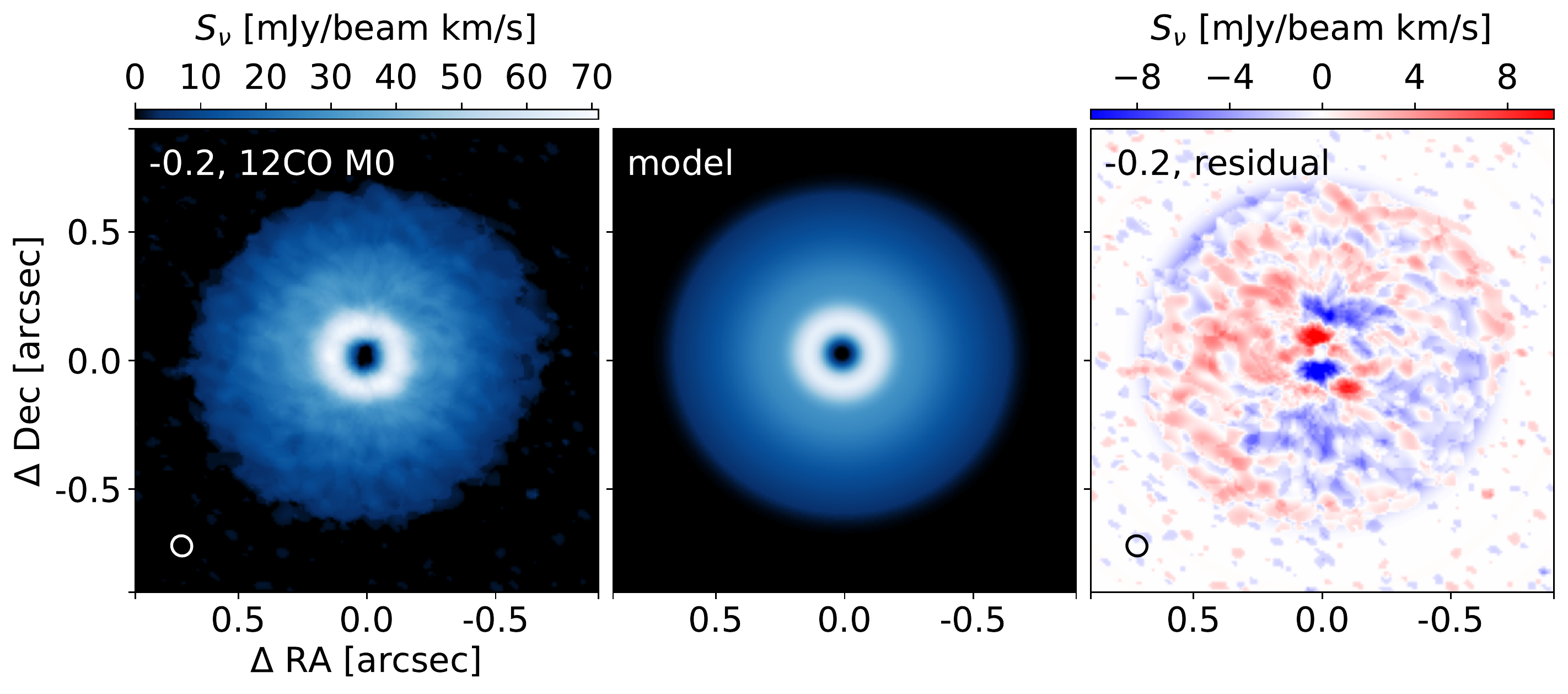}\\
        \includegraphics[width=15cm]{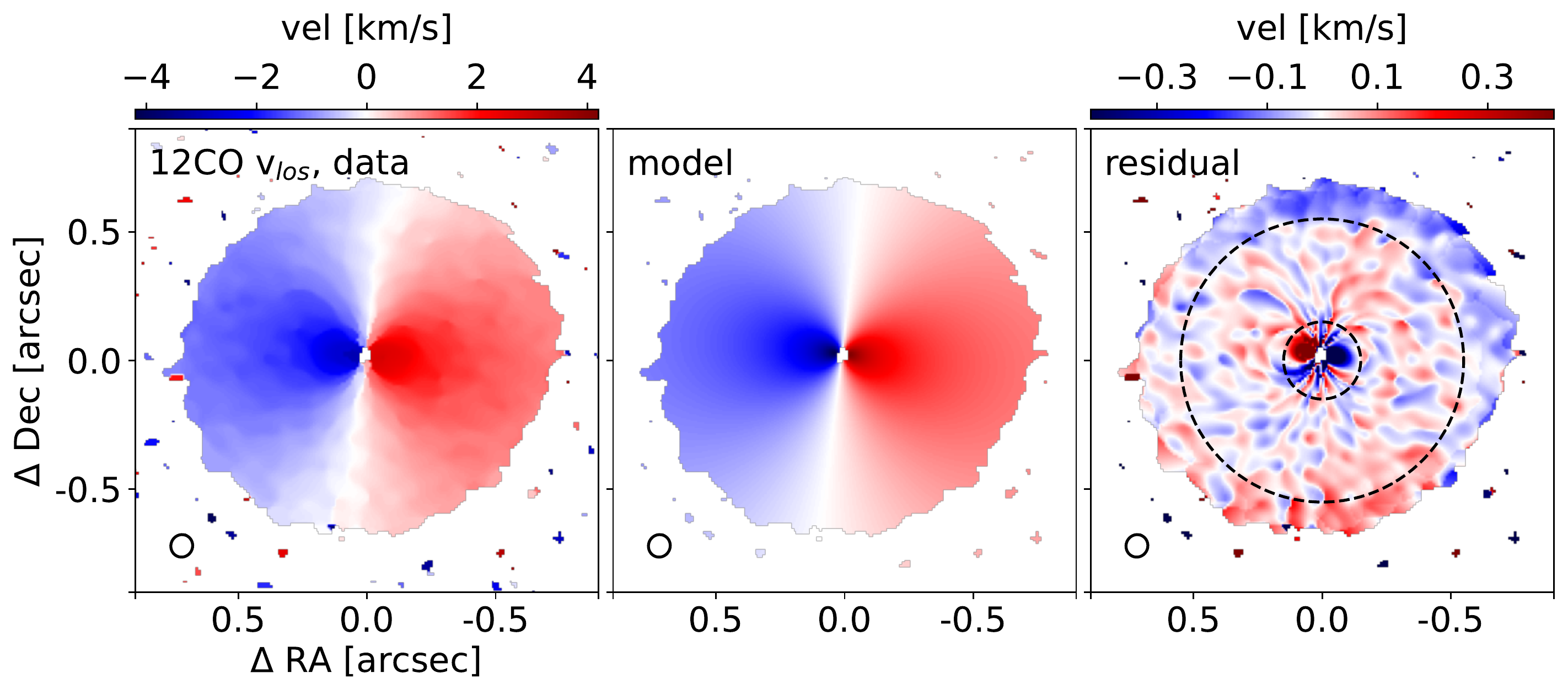}\\
        \vspace{-0.1cm}
   \caption{CS\,Cha gas emission and kinematics. \textbf{Upper row}: $^{12}$CO moment 0, the best model using the geometry recovered from the kinematic fit, and the residuals. Scale bar represents 20\,au at the distance of the source. \textbf{Lower row}: $^{12}$CO peak velocity in the line of sight, with the best model calculated with the parameters from Table~\ref{tab:eddy_results}, and the residuals. The dashed line shows the mask used for the fit.}
   \label{fig:moms}
\end{figure*}

We generated images for the $^{12}$CO emission from CS\,Cha with different robust parameters to check the emission at a high angular resolution, but also to check the extended structure with high S/N. The channel maps were all generated with the same velocity channels, and their only difference is the robust parameter used. A gallery of the channels generated with robust 0.0 is shown in the appendix (Fig.~\ref{fig:channel_maps}).

The $^{12}$CO emission (shown in Figs.~\ref{fig:obs_images} and \ref{fig:moms}) appears depleted in the central region of the cavity, and the brightness peak is located at 128\,mas (or 21.6\,au), which is closer to the center of the disk compared to the dust continuum radial profile, which peaks at 34.6\,au. 
The profile recovered from the different moment 0 images consistently show the brightness peak at the same radial location. By using the inclination from the continuum fit, plus the vertical structure and PA traced by a kinematic fit with the \texttt{eddy} package \citep{eddy} (as described in the following Section~\ref{sec:kinematics}, and summarized in Table~\ref{tab:eddy_results}), we deprojected the $^{12}$CO Moment 0 image and used it to calculate an azimuthally averaged surface brightness profile (shown in Fig.~\ref{fig:moms}).

We subtracted the azimuthally averaged surface profile from the $^{12}$CO moment 0, to search for asymmetries. The moment 0 is preferred over the peak intensity map as the later is more affected by the beam size and geometry of the disk, thus creating overbrightness regions along the major axis which are not of physical origin. When an azimuthally symmetric model is subtracted from the moment 0, the disk shows residuals on extended and compact scales, with a typical contrast between the emission and model on the order of $<15\%$. The brightness temperature of the $^{12}$CO moment 0 reaches about 120\,K at the radial profile peak and decreases towards about 10\,K in the outer edge. Due to this temperature range, the emission is possibly more optically thick in some regions than in others and, thus, its brightness traces a combination of temperature and gas density variations at the disk surface layers. These residuals may originate from a combination of small-scale height variations, disk eccentricity and dynamical perturbations, and none are included in the azimuthally averaged surface profile.

\subsubsection{Kinematics of the $^{12}$CO emission}\label{sec:kinematics}

We calculated the velocity map of the $^{12}$CO by using the package \texttt{bettermoments}, which fits a quadratic function to each pixel over the channel maps cube, allowing us to obtain the velocity corresponding to the peak emission with sub-channel velocity resolution. The velocity map used in the kinematic analysis is shown in Figs.~\ref{fig:obs_images} and~\ref{fig:moms}. Additionally, a careful analysis of the channel maps in Fig.~\ref{fig:channel_maps} allows us to confirm that the southern side of the disk is closer towards us, and so the disk is rotating counter-clockwise from the observers' point of view. This coincides with the projected direction of the proposed orbits for CS\,Cha\,B \citep{ginski2018}, however, its orbital plane has not been accurately constrained and it might not necessarily be coplanar to CS\,Cha.

As discussed in the Section~\ref{sec:cont_morph}, with the disk being so close to face-on, there are correlations that are difficult to disentangle without fixing some geometric parameters. In the case of the $^{12}$CO kinematic image, there is a strong correlation between the total mass of the central stars $M_{\text{total}}$, the inclination of the disk (inc), and the surface layer geometry from where the $^{12}$CO is being detected, which we describe as a function of the radius from the center of the disk ($h_{\text{CO}}(r)$). Due to the low inclination of the disk, a variable that is mostly independent from the previous unknown parameters is the PA of the disk, and so it is the first value that we constrain. 

We used the  \texttt{eddy} package to fit the $^{12}$CO kinematic map under the assumption of flat disk, to avoid introducing additional free parameters while the inclination and \mtot\, are still not constrained. We run a MCMC with uniform prior over the six free parameters that include the center of the disk ($x_0$, $y_0$), the disk geometry (inc, PA), the binaries mass (\mtot), and velocity at the local standard of reference (VLSR), and we recovered a value of PA$=262.6\pm0.1$\,deg, which is consistent for kinematic maps generated from different robust parameters. This value is higher than $180$\,deg since the convention used in this kinematic fitting is that the PA is aligned with the red-shifted part of the disk. If we follow the dust continuum emission convention of measuring the PA as the angle between the north and the semi-major axis to the east, we obtain PA$=82.6\,$deg (this includes an assumption of a flat dust continuum disk). This value is used in the uv modeling of the dust continuum (shown in Section~\ref{sec:cont_morph}), from where we find an inclination of the midplane of inc$=17.86$\,deg, which we assume to be the inclination for the $^{12}$CO emission.

By having the inclination fixed, the degeneracy of the value for \mtot\, is reduced, enabling us to include as free parameters the description for the vertical height ($z_{\text{CO}}(r)$) of the emitting surface layer. We perform this new fit under the assumption of a single power law, following $z_{\text{CO}} \,=\, z_0 \, \cdot \, r^\psi$, where the free parameters are the pair ($z_0, \psi$), and it is only a function of the distance to the disk center $r$. The Keplerian velocity is calculated by including the scale of the height $^{12}$CO in the distance to the center of the disk, based on the following:
\begin{equation}
    v_{\text{kep}}(r, z) \,=\, \sqrt{\frac{G\,M_{\text{total}}\,r^2 }{(r^2+z}^2)^{3/2}}\text{.}\\
    \label{eq:kep_inclh}
\end{equation}

The best parameters obtained after running a MCMC optimizer with the new model are listed in Table~\ref{tab:eddy_results}, where we recover the central mass of the stars: \mtot$\approx1.91\,M_\odot$. The kinematic image of the best model, and the residuals, are shown in the bottom-middle and bottom-right panels of Fig.~\ref{fig:moms}, respectively, where the mask used to fit the velocity map is shown: an annulus with inner radius of 0.15'' and outer radius of $0.65''$. Given that our model does not includes eccentricity, the inclusion or exclusion of different regions of the disk can change the position of the centroid, which, in turn, also affects the best fit parameters. Depending on the masked region used to fit the velocity map, the mass of the central object can shift between $1.86-1.91\,M_\odot$, due to changes in the position of the center. The non-eccentric kinematic model is also the reason for which the values of $\delta_{\rm{RA}}$ and $\delta_{\rm{Dec}}$ do not match between the dust continuum and $^{12}$CO fits.

In principle, the eccentricity is expected to decrease for regions that are located farther away from the binaries. Fitting those regions with a kinematic model should therefore lead to a better determination of the disk barycenter position. In CS\,Cha, however, there are two issues with including the outer-regions in the velocity fit: i) the S/N is decreased towards the outer edge of the disk, thus not allowing us to distinguish between the emission from the front-side and back-side of the disk; and ii) the line following the zeroth velocity at different radius (in our line of sight), known as the line of nodes, is curved in the outer regions of the disk (as can be better seen in the left panel of Fig.~\ref{fig:moms} and channel map 3.65 in Fig.~\ref{fig:channel_maps}). 
The mechanisms driving the velocity residuals inside and outside of the mask are still a subject to be studied. In the disk cavity, the residuals could be a combination of eccentric gas flow due to the binaries, and also radial flows of material flowing from the main ring towards the binaries, as described in \citet{rosenfeld2014}. As for the outer disk, the residual velocities could show a combination of eccentricity (which is not accounted in a circular model), and tidal influence from the companion CS\,Cha\,B. Such tidal interaction has been observed in other disks in multiple-stellar systems, such as AS\,205 and RW\,Aur \citep[][respectively]{kurtovic2018, rodriguez2018}.

\begin{table}[t]
\centering
\begin{tabular}{ c|c|c } 
  \hline
  \hline
\noalign{\smallskip}
Parameter            & Best fit                  & units \\
\noalign{\smallskip}
  \hline
\noalign{\smallskip}
 $\delta_{\rm{RA}}$  & $6.9 \,\pm\, 0.1$         & mas \\
 $\delta_{\rm{Dec}}$ & $28.8 \,\pm\, 0.4$        & mas \\
 inc                 & $17.86$ fixed             & deg \\
 PA                  & $263.1 \,\pm\, 0.2$       & deg \\
\noalign{\smallskip}
  \hline
\noalign{\smallskip}
 $M_\star$           & $1.911 \,\pm\, 0.002$     & $M_\odot$ \\
 VLSR                & $3670.1 \,\pm\, 0.4$      & m\,s$^{-1}$ \\
  \hline
\noalign{\smallskip}
 $z_0$               & $0.024 \,\pm\, 0.002$      & arcsec \\
 $\psi$              & $0.033 \,\pm\, 0.022$      & - \\
\noalign{\smallskip}
  \hline
  \hline
\end{tabular}
\caption{Best parameters from the kinematic fitting with \texttt{eddy} to the image generated with a robust parameter of 0.0. The vertical profile is calculated following $z(r) \,=\, z_0 \, \cdot \, r^\psi$.}
\label{tab:eddy_results}
\end{table}

%%%%%%%%%%%%%%%%%%%%%%%%%%%%%%%%%%%%%%%%%%%%%%%%%%%%%%%%%%%%%%%%%%%%%%%%%%%%%%%%

\section{Hydrodynamical simulations \label{sec:hydro_sim}}

We compared our ALMA observations with hydrodynamical simulations of circumbinary disks, to test the general conditions that could generate the observable characteristics of CS\,Cha. Our main focus was the comparison between the cavity and ring properties in circumbinary disks that do and those that do not host a single Saturn-like planet, as we describe in the following subsections.

\subsection{Setup: Circumbinary disk with FARGO3D}

%to measure the difference between a circumbinary disk in CSCha with and without a planet
We ran our simulations in a modified version of the FARGO3D code \citep{benitez2016} used in \citet{thun2018} to simulate a 2D-hydrodynamical model of a circumbinary disk with the stellar properties of CS\,Cha. 
As in \citet{thun2018} and \citet{penzlin2021}, we simulated the disk with a cylindrical grid, starting at an inner radius of $1\,a_{\text{bin}}$ up to $40\,a_{\text{bin}}$ (with $a_{\text{bin}}$ the binary separation), with 684 logarithmically spaced radial cells and 1168 azimuthal cells, which is twice the azimuthal resolution used in the previously mentioned works. This higher resolution is applied to ensure convergence when the planet is included, and the results with half resolution are consistent with the ones shown in the following sections. 
The total binary mass was obtained from our \texttt{eddy} fit ({\mtot} $=1.91\,M_\odot$), and the mass ratio used for the binaries is $q=0.7$, calculated in Ginski et al. (in prep.) by using the R-band and I-band magnitudes of {\footnotesize SPHERE ZIMPOL} observations \citep{beuzit2019, schmid2018}.  

For the surface density, we used a radially dependent profile given by $\Sigma(r) = f_{\text{gap}} \cdot \Sigma_0 \cdot r^{-3/2}$, where $r$ is the radial distance from the disk center, $\Sigma_0$ is chosen such that the disk mass is $0.01\,$\mtot, and $f_{\text{gap}}$ is an exponential function that depletes the density profile inside $2.5\,a_{\text{bin}}$, therefore speeding up the number of orbits needed to reach the steady state.  
We calculate $f_{\text{gap}}$ as in \citet{thun2017}, by using $f_{\text{gap}} = \left( 1 + \exp{[- (r-2.5\,a_{\text{bin}}) / (0.25\,a_{\text{bin}}) ]} \right)^{-1}$. 
All our simulations have fixed $\alpha$ viscosity parameter of $\alpha=10^{-4}$ \citep{shakura1973} for all radii. 

Our simulations use a locally isothermal equation of state for the gas, which allows for a faster convergence compared to a viscous heated radiative disk.
In the latter, the steady state of the gas is comparable to the isothermal setup for a constant disk aspect ratio $h/r$, but it can take over $100\,000$ orbits to be reached \citep{kley2019}. We set our binaries such that they are only sensitive to each other (and not to the disk around them). 

Due to the long period of their orbit \citep[at least 2482\,days, ][]{guenther2007}, the separation of the components and their eccentricity is only constrained to be within a certain parameter range ($a_{\text{bin}}<7.5\,$au and $e_{\text{bin}}\lesssim0.5$, Ginski et al. in prep.). Given that the simulations can be run with normalized units, the uncertainty in the binary separation can be circumvented by treating distances in terms of the binary separation, but to account for the possible binary eccentricities, we need to run simulations with different values. 
Therefore, we ran three different binary eccentricities setups: $e_{\text{bin}}=[0.15,\,0.25, \,0.35]$, and we let them evolve for $20\,000$ binary orbits to reach the steady state. 
As the circumbinary disks have minimal eccentricity for $e_{\text{bin}}\approx0.15$, we sampled the allowed eccentricity range for $e_{\text{bin}}$ with increasing eccentricities starting from the smallest \citep{thun2018, kley2019}. 
Additionally, previous works have shown that different aspect ratios can have an impact in the disk gas morphology \citep{thun2018, tiede2020, penzlin2021}; therefore, we ran each binary eccentricity with 2 aspect ratios: $h/r=0.03$ and $h/r=0.05$. 

After 20\,000 orbits, we introduced a single planet at a distance of $6\,a_{\text{bin}}$ and we let it migrate inward to its equilibrium orbit. From \citet{kley2019}, we know that the planet ability to open a gap determines its evolution. Planets that are able to open a gap can separate the outer disk from the inner disk, effectively shielding the outer disk from the binaries action, lowering the eccentricity of these regions. Since we are running simulations in a low viscosity scenario, we decided to use a giant planet of low mass $M_p=1\,M_{\text{Saturn}}$, which is consistent with the planets detected in P-type orbits \citep{penzlin2021}. 

Previous studies have also found that more massive planets, such as 1$M_{Jup}$, are prone to more unstable orbits and have a higher likelihood of getting excited into a larger distance orbit or even of getting ejected from the system \citep{pierens2008}.

After introducing the planet, we ran each simulation for another $100\,000$ binary orbits, which is $50\,000$ orbits after the convergence of five out of six of our migrating planets. For comparison, we also kept running the simulations without a planet for $100\,000$ additional binary orbits. This leaves us with 12 simulations when taking into account all binary eccentricities, disk aspect ratios, and planet presence. A summary of the setups is found in Table~\ref{tab:sim_results}.

\subsection{Disk evolution with no planet}\label{sec:disk_evo_nopl}

\begin{figure*}[t]
 \centering
        \includegraphics[width=18cm]{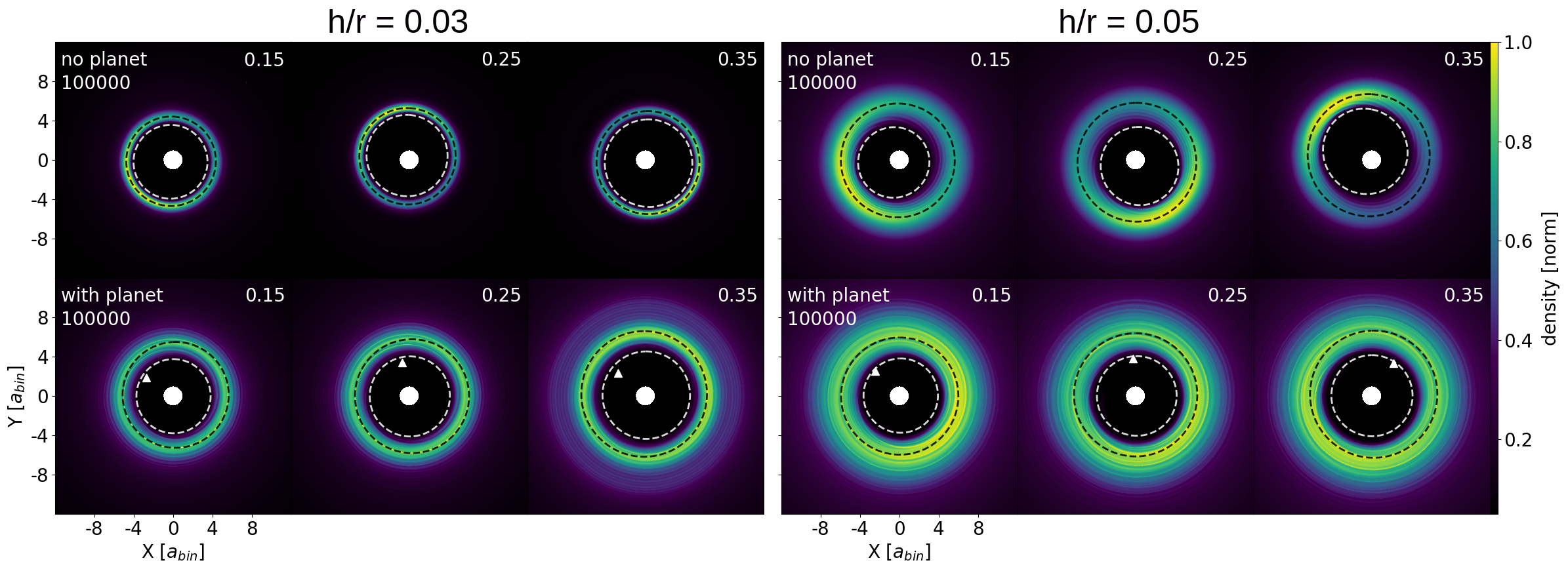}\\
        \vspace{-0.1cm}
   \caption{Gas surface density after 100\,000 binary orbits in each setup. Distance is in binary separations and the color scale is normalized to the maximum of each image. 
   Panels on the left and right show the setups with $h/r=0.03$ and $h/r=0.05$ respectively. In each panel, the columns show the setups with the same binary eccentricity, being 0.15, 0.25, and 0.35 from left to right. The upper row of each panel contains the setups with no planet, and the lower row the setups with planet. A white dashed line shows the best cavity fit, while the black dashed line shows the best peak ring fit. A white triangle is used to show the position of the planet.}
   \label{fig:last_output}
\end{figure*}

\begin{table*}[t]
\centering
\begin{tabular}{ c||c|c|c|c|c|c||c|c|c|c|c|c } 
  \hline
  \hline
\noalign{\smallskip}
 & \multicolumn{6}{c||}{$h/r=0.03$} & \multicolumn{6}{c}{$h/r=0.05$} \\ 
  \hline
 & \multicolumn{3}{c|}{no planet} & \multicolumn{3}{c||}{with planet} & \multicolumn{3}{c|}{no planet} & \multicolumn{3}{c}{with planet}  \\ 
  \hline
setup & 0.15 & 0.25 & 0.35 & 0.15 & 0.25 & 0.35 & 0.15 & 0.25 & 0.35 & 0.15 & 0.25 & 0.35 \\
\noalign{\smallskip}
  \hline
  \hline
\noalign{\smallskip}
$e_{\text{cav}}$  & 0.089 & 0.093 & 0.122 & 0.015 & 0.032 & 0.027 & 0.153 & 0.196 & 0.206 & 0.028 & 0.043 & 0.044 \\
${a}_{\text{cav}}$  & 3.87  & 4.25  & 4.59  & 3.79  & 4.09  & 4.47  & 3.76  & 4.13  & 4.52  & 3.72  & 3.90  & 3.99 \\
  \hline
$e_{\text{peak}}$ & 0.061 & 0.069 & 0.094 & \textcolor{teal}{\textbf{0.049}} & \textcolor{teal}{\textbf{0.046}} & \textcolor{teal}{\textbf{0.034}} & \textcolor{teal}{\textbf{0.028}} & \textcolor{teal}{\textbf{0.056}} & 0.079 & \textcolor{teal}{\textbf{0.049}} & \textcolor{teal}{\textbf{0.045}} & \textcolor{teal}{\textbf{0.037}} \\
${a}_{\text{peak}}$ & 4.55  & 4.91  & 5.24  & 5.39  & 5.78  & 6.39  & 5.79  & 6.05  & 6.22  & 6.06  & 6.27  & 6.57 \\
$q_{\text{peak}}$ & 1.43  & 1.44  & 1.60  & \textcolor{teal}{\textbf{1.16}}  & \textcolor{teal}{\textbf{1.16}}  & \textcolor{teal}{\textbf{1.10}}  & 1.40  & 1.60  & 1.76  & \textcolor{teal}{\textbf{1.11}}  & \textcolor{teal}{\textbf{1.18}}  & \textcolor{teal}{\textbf{1.17}}  \\
  \hline
$e_{\text{pl}}$   &  &  &  & 0.019 & 0.020 & 0.024 &  &  &  & 0.016 & 0.017 & 0.020 \\
$a_{\text{pl}}$   &  &  &  & 3.34  & 3.55  & 3.74  &  &  &  & 3.58  & 3.83  & 3.99  \\
\noalign{\smallskip}
  \hline
  \hline
\end{tabular}
\caption{Eccentricity and semi-major axis of the cavity edge ($e_{\text{cav}}$, $a_{\text{cav}}$), peak density ($e_{\text{peak}}$, $a_{\text{peak}}$), and planetary orbit ($e_{\text{pl}}$, $a_{\text{pl}}$). Values were calculated by taking the median of the last 1000 binary orbits. The ratio between the brightest and dimmest part of the ring peak is shown as $q_{\text{peak}}$. The highlighted $e_{\text{peak}}$ and $q_{\text{peak}}$ are the values consistent with the observations, as described in Sect.~\ref{sec:disc_planet}.}
\label{tab:sim_results}
\end{table*}

In the absence of a planet, the disk cavity quickly becomes eccentric, with the size of the cavity and its precession velocity being dependent on the eccentricity of the binaries ($e_{\text{bin}}$) and the disk aspect ratio ($h/r$). In order to measure the cavity properties, we trace the cavity boundary by searching for the radial position at which the density reaches $10\%$ of the peak density, and we repeat for every azimuthal element of the gas density image, thus obtaining 1168 radial positions for each time step. This $10\%$ threshold is chosen to avoid the streamers of material that flow from the circumbinary ring onto the binaries. We fit these points with an eccentric orbit by using the function \texttt{curve\_fit} from the Python package \texttt{scipy.optimize} \citep{scipy2020}. The best fit allows us to recover the eccentricity of the cavity ($e_{\text{cav}}$), the semi-major axis ($a_{\text{cav}}$), and the argument of the periastron ($\omega_{\text{cav}}$), which is used to trace the cavity precession.

We show the cavity boundary fit for the binary orbit 100\,000 with a white dashed line in Fig.~\ref{fig:last_output}, and the median value of $a_{\text{cav}}$ and $e_{\text{cav}}$ the last 1000 binary orbits is shown in Table~\ref{tab:sim_results}. Alternatively, another approach to recover the eccentricity information of the disk is through the eccentricity vector, which uses the kinematic information and returns the eccentricity of a gas parcel. We find consistent results between both methods (calculating from density compared to eccentricity vector), and we decided to go with the density-based estimation to be consistent with our uv modeling approach to recover eccentricity.

\begin{comment}
\begin{figure*}[t]
 \centering
        \includegraphics[width=18cm]{figures/FARGO/cavity_properties_hr3.png}\\
        \includegraphics[width=18cm]{figures/FARGO/cavity_properties_hr5.png}\\
        \includegraphics[width=18cm]{figures/FARGO/peak_properties_hr3.png}\\
        \includegraphics[width=18cm]{figures/FARGO/peak_properties_hr5.png}\\
        \vspace{-0.1cm}
   \caption{Eccentricity and semi-major axis for the cavity ($\text{e}_{\text{cav}}, \text{a}_{\text{cav}}$) and the ring peak ($\text{e}_{\text{peak}}, \text{a}_{\text{peak}}$) as a function of time in binary orbits, shown blue-like and red-like colors respectively. Each panel-pair contains the simulations of the 3 different eccentricity setups for the binaries, and a single aspect ratio, displayed in the upper left corner together with the line-style legend. The setups without planets are shown in darker color, while setups with planets are softer in color and with thinner lines, to show the fast variations as a function of time.}
   \label{fig:cavity_prop}
\end{figure*}
\end{comment}

\begin{figure*}[t]
 \centering
        \includegraphics[width=18cm]{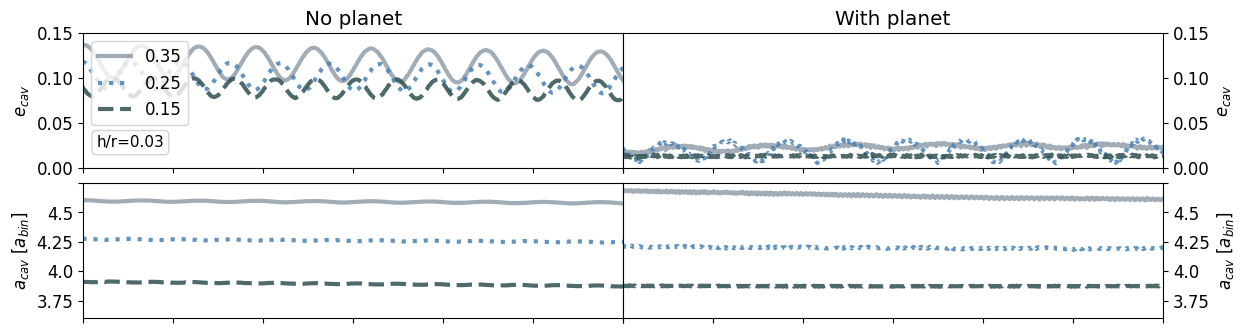}\\
        \includegraphics[width=18cm]{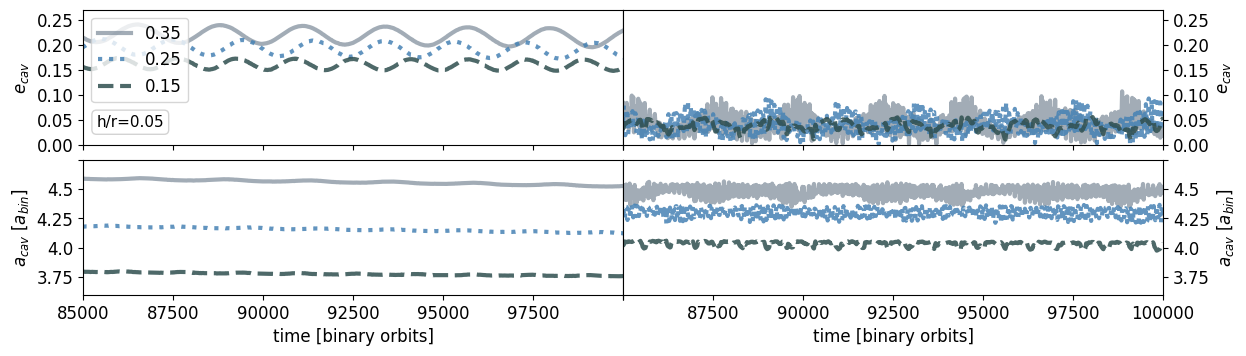}\\
        \vspace{-0.1cm}
   \caption{Eccentricity and semi-major axis for the cavity ($\text{e}_{\text{cav}}, \text{a}_{\text{cav}}$) as a function of time in binary orbits. Each panel contains the simulations of the 3 different $e_{\text{bin}}$ for the binaries, and a single aspect ratio, displayed in the left-side together with the line-style legend.} 
   \label{fig:cavity_prop}
\end{figure*}

\begin{figure*}[t]
 \centering
        \includegraphics[width=18cm]{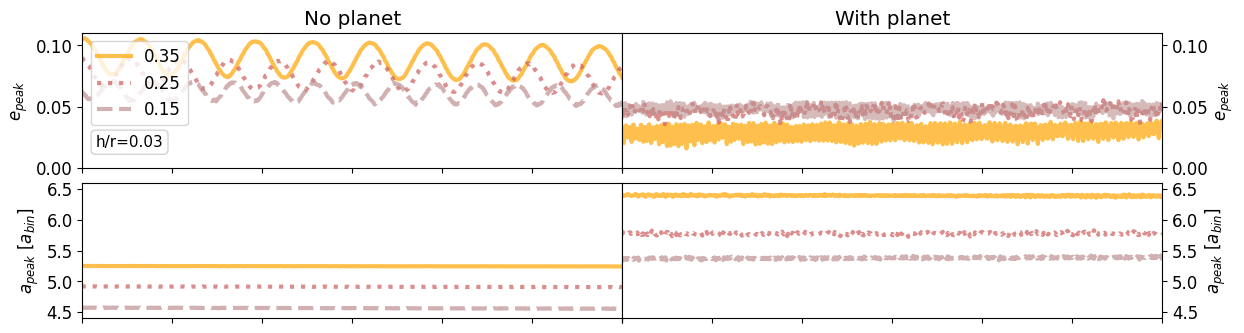}\\
        \includegraphics[width=18cm]{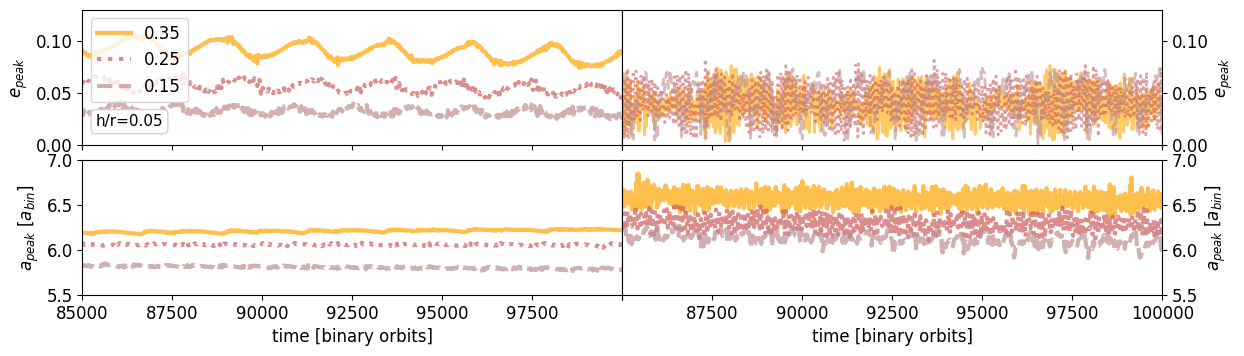}\\
        \vspace{-0.1cm}
   \caption{Eccentricity and semi-major axis for the peak density ring ($\text{e}_{\text{peak}}, \text{a}_{\text{peak}}$) as a function of time in binary orbits. Each panel contains the simulations of the three different $e_{\text{bin}}$ values for the binaries and a single aspect ratio, displayed in the left-side together with the line-style legend.}
   \label{fig:peak_prop}
\end{figure*}

As expected, the smallest $e_{\text{cav}}$ and $a_{\text{cav}}$ are obtained for the binaries with eccentricity of 0.15, which at the end of the simulation have a semi-major axis $a_{\text{cav}}<4\,a_{\text{bin}}$ in both aspect ratio setups. The biggest cavity sizes of all setups are found for the binaries with eccentricity of 0.35, with a final cavity size $a_{\text{cav}}>4.5\,a_{\text{bin}}$. Overall, the difference between smallest and biggest cavities is only about $15\%$. We also find that neither $e_{\text{cav}}$ nor $a_{\text{cav}}$ are constant in time, as they have an oscillatory behavior with shorter period for smaller binary eccentricity, shown in Fig.~\ref{fig:cavity_prop}. A quick analysis with a periodogram allowed us to find that the oscillations period in $e_{\text{cav}}$ and $a_{\text{cav}}$ are almost identical to the precession period of $\omega_{\text{cav}}$, which ranges between 1100 to 2300\,$T_{\text{bin}}$ depending on the binary eccentricity and disk aspect ratio. Considering a binary period of 7\,yr for CS\,Cha, a complete precession of the cavity would be observable over a period of at least 7700\,yr, considering the shortest precession period of our simulations. 

%Since our simulations are locally isothermal, the position of the gas pressure maxima coincides with the gas density maxima.
We also traced the radial positions at which the gas density is highest for each azimuthal element (a peak gas density ring), and we fit an eccentric orbit in the same way it was done for the cavity, allowing the recovery of the eccentricity $e_{\text{peak}}$, semi-major axis $a_{\text{peak}}$, and argument of the periastron $\omega_{\text{peak}}$. The best fit to the peak positions is shown in Fig.~\ref{fig:last_output} with black dashed lines for the binary orbit 100\,000, and the median values for the last 1000 binary orbits for all setups is shown in Table~\ref{tab:sim_results}. In the absence of a planet the values for $e_{\text{peak}}$ can go from 0.03 to almost 0.1, while $a_{\text{peak}}$ ranges between 4.6 and 6.2\,$a_{\text{bin}}$. Similar to the cavity, the peak density ring also shows an oscillatory behavior on its parameters, which is shown in Fig.~\ref{fig:peak_prop} for the last 15\,000 orbits (in reddish colors). 

Another significant feature that can be drawn from the simulations is the azimuthal density variation along the density peaks, which can be better appreciated in Fig.~\ref{fig:last_output}. Due to the eccentricity of the disks, there is an over-density at the location of the apoastron of the orbits, which is a product of the slower orbital velocities at that location compared to the periastron orbital velocity. We measure the contrast between the highest density and lower density along the peak density ring by calculating $q_{\text{peak}}=\rho_{\max}/\rho_{\min}$, where $\rho$ is the gas density. This value is reported in Table~\ref{tab:sim_results}, and we find that the binaries with 0.15 of eccentricity have the least asymmetric rings for the no planet setups, with an excess of at least $40\%$ between maximum and minimum density.

\subsection{Disk evolution with a Saturn-like planet}

\begin{figure*}[t]
 \centering
        \includegraphics[width=15cm]{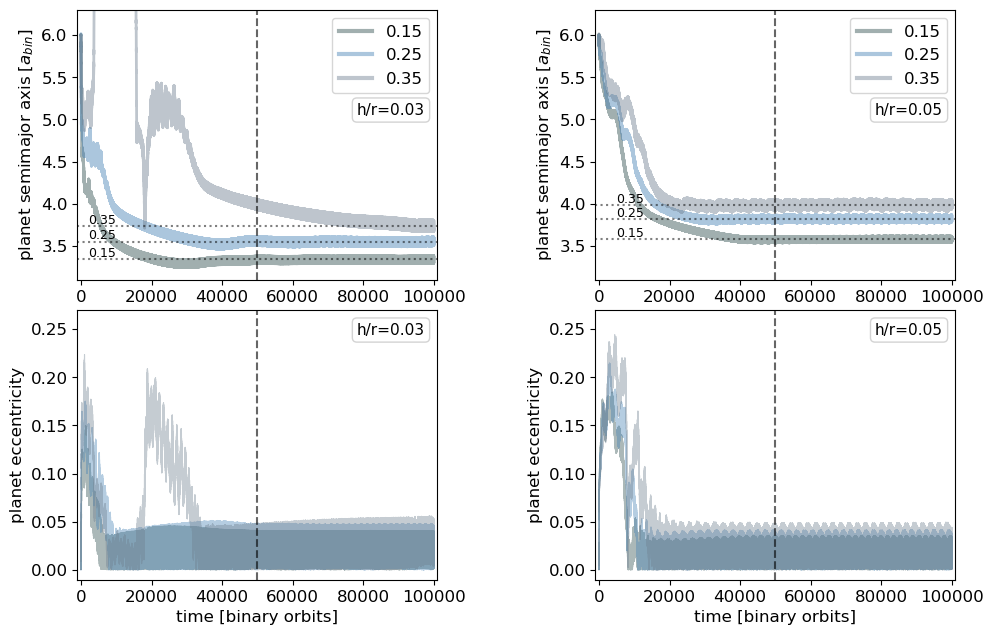}\\
        \vspace{-0.1cm}
   \caption{Planet's semi-major axis and eccentricity as a function of time (in binary orbits) in the upper and lower panel row, respectively. The upper right number in each plot indicates the eccentricity of the binaries. The dashed vertical curve indicates the position of the 50\,000 binary orbits, after which the planet has converged to its equilibrium orbit. The median semi-major axis for the planet orbit after convergence is indicated with a dotted line.}
   \label{fig:planets_figure}
\end{figure*}

The planet starts migrating inwards and carving a gap as soon as it is introduced into the simulation. The evolution of the planet's eccentricity and semi-major axis is shown in Fig.~\ref{fig:planets_figure}. Depending on the binary eccentricity, the planet takes different times to converge to its steady orbit, and the longest time for convergence is obtained for the planet around binaries of eccentricity $e_{\text{bin}}=0.15$. After 50\,000 binary orbits, the planet in most of the simulations has converged to its steady semi-major axis ($a_{\text{pl}}$), which ranges between 3.3 to 4.0\,$a_{\text{bin}}$, depending on the binary eccentricity and disk aspect ratio. 
The only planet that takes more than 50\,000 binary orbits to converge to its final position is the planet in the setup $e_{\text{bin}}=0.35$ with $h/r=0.03$, as shown in the left panel of Fig.~\ref{fig:planets_figure}. As the eccentricity of the binaries is increased, the instability region is pushed farther away, thus the initial position of this planet was more unstable than the others. The planet is initially pushed to a farther orbit, before it starts migrating inwards as the others. Examples of this behavior are also seen in \citet{penzlin2021}, where some of the planets would even get ejected from the system depending on the binary mass ratio, eccentricity and disk aspect ratio, for the same initial planet position. As this planet jumps into a higher orbit, it creates a secondary ring outside the main ring excited by the binaries, with a gap between them located roughly at 8\,$a_{\text{bin}}$. After the planet has migrated inwards into the cavity, the secondary ring remains as a stable structure until the end of our simulations.

The eccentricity of the planet orbit $e_{\text{pl}}$ is not constant with time, and it oscillates between 0 and 0.05 in both aspect ratio setups (as shown in the lower panel of Fig.~\ref{fig:planets_figure}). By analyzing the periodogram of $e_{\text{pl}}$ (after $a_{\text{pl}}$ convergence), we find the oscillation period to be consistent with the cavity oscillations periods, in agreement with the findings of \citet{penzlin2019}, where multiple planets were considered.

The planets modify the structure of the cavity and the overall disk eccentricity. To quantify the difference between the setups with and without planets, we calculated the cavity properties and peak density ring properties following the same procedures explained in Section~\ref{sec:disk_evo_nopl}. The planet presence considerably decreases $e_{\text{cav}}$, as shown in Fig.~\ref{fig:cavity_prop}, where the highest amplitude variations do not reach the minimum $e_{\text{cav}}$ from the no planet setups, independently from the $e_{\text{bin}}$ and disk aspect ratio. This oscillations are consistently confined to the range between (0.0, 0.1) for the aspect ratio $h/r=0.05$, and (0.0, 0.03) for the aspect ratio $h/r=0.03$. As in the "no-planet" simulations, the period of the oscillations in $e_{\text{cav}}$ and $a_{\text{cav}}$ are consistent with the cavity precession period, and they coincide with the oscillation period of $e_{\text{pl}}$. 

The decrease in eccentricity due to the planet's presence is extended towards the whole disk. The only case where peak ring eccentricities become comparable between setups with or without a planet is for the aspect ratio $h/r=0.05$, where the setups without planet and $e_{\text{bin}}=0.15$ and $0.25$ have $e_{\text{peak}}$ in the same eccentricity range of with planet setups (as shown bottom panel-pair in Fig.~\ref{fig:peak_prop}). The overall decrease in eccentricity contributes to a decrease in the density asymmetry along the peak density ring, as the gas spends similar amount of time in each azimuthal element. Considering all our simulations, we find the gas density profile to be between three to seven times more axisymmetric when the planet is present, as shown in Table~\ref{tab:sim_results}.

%%%%%%%%%%%%%%%%%%%%%%%%%%%%%%%%%%%%%%%%%%%%%%%%%%%%%%%%%%%%%%%%%%%%%%%%%%%%%%%%

\section{Discussion \label{sec:discussion}}

\subsection{Non-detetion of CS\,Cha\,B }

Despite the strong evidence of a highly inclined disk around CS\,Cha\,B \citep[high polarization fraction, optical and NIR attenuation, and accretion rate][]{ginski2018, haffert2020}, our observations are unable to detect such material at 0.87\,mm wavelength. For comparison, the $35.4\mu$Jy limit is three times fainter than the detected flux from PDS70c \citep{benisty2021} or the free floating planet OTS\,44 \citep{bayo2017} -- and it is even lower than the upper limits found for protolunar disk fluxes of directly imaged exoplanets \citep{perez2019}.

The non-detection of CO towards CS\,Cha\,B suggests that its emission is either being blocked, or that its CO emitting layer is very compact (or a combination of both). Disks around M-dwarf stars are expected to be smaller compared to disks around Sun-like stars \citep{andrews2013,tripathi2017,hendler2020}, and the tidal interaction of CS\,Cha\,B with the main CS\,Cha system could have further truncated its size \citep{bate2018,cuello2019,manara2019}. Observations at shorter wavelengths (such as ALMA bands 8-10 or JWST instruments MIRI and NIRcam) are needed to fully understand the circumstellar environment of CS\,Cha\,B, by connecting the non detection in $0.87$\,mm to the NIR observations.

Alternatively, CS\,Cha\,B could also be a young source located in the background of the Cha I cloud. In such scenario, a disk-less star (or a very small disk) would explain the non-detection at mm wavelengths, while the light would be additionally obscured and polarized by the environment. The H$\alpha$ emission could be a contribution from accretion and chromospheric activity \citep[e.g. PZ Tel B,][]{barcucci2019}, and the apparently common proper motion would be due to both systems being on the same cloud. A longer time baseline on the sources astrometry could clarify whether the sources are indeed gravitationally bounded or whether their apparent proximity and similar proper motion is only a temporary coincidence.

\subsection{A Saturn-like planet is consistent with the morphology of the CS\,Cha disk}\label{sec:disc_planet}

% eccentricity and azimuthal symmetry are related
Two main properties distinguish the simulated disks that host a Saturn-like planet to the ones that do not: the disk eccentricity and azimuthal density symmetry (or azimuthal contrast). These properties are not independent from each other. In fact, binary disks that do not host a planet have cavities that are consistently more eccentric compared to the planet hosting disks, and a similar behavior is seen for the ring eccentricity. The increased eccentricity produces a higher difference between the orbital velocity at apoastron and periastron, which contributes to the azimuthal asymmetry. 

Our simulations are all locally isothermal, and so the gas density maxima will coincide with the gas pressure maxima, where the dust is expected to be trapped more efficiently. As a first approximation, we can compare the eccentricity of the gas density peaks from the simulations to the eccentricity of the dust continuum peaks from our uv modeling. This assumes that the trapped dust in the pressure bump has the same eccentricity than the gas \citep[as in][]{ataiee2013}. To quantify how coupled are the dust particles to the gas, we check the Stokes number of the 1mm-sized particles at the location of the density peak in each simulation, by following the formulas presented in \citet{birnstiel2016}. We find typical values ranging between 0.015 and 0.03 (assuming a volume density of the particles of $1.2\,$g\,cm$^{-3}$). Particles with such Stokes numbers are prompt to be trapped in pressure maxima, in particular when the disk viscosity is low \citep{pinilla2012, birnstiel2013, dejuanovelar2016}. Hence, in the framework of our simulations ($\alpha=10^{-4}$), the assumption of the eccentricity of the gas density peak to be equal to the eccentricity of the dust continuum peak is valid. 

The dust continuum observations are better suited than the $^{12}$CO images to be compared with the simulations because the 0.87mm continuum traces the dust density at the midplane (in the optically thin approximation, with a constant temperature at different radii), while the optically thick $^{12}$CO traces temperature in the disk surface layers. Therefore, the peak emission in the $^{12}$CO moment 0 is showing regions of high temperature, and it does not trace gas surface density. In the best parametric model (Model 2e, see Section~\ref{sec:cont_morph} and Table~\ref{tab:mcmc_results}), we find that the eccentricity of the component $g_0$, which describes the ring peak (see schematic in Fig.~\ref{fig:schematic_profiles}), is $e_{g_0}=0.039$. In the following discussion, we consider this value as reference to be compared with $e_{\text{peak}}$.

% Eccentricity is consistent with several setups
In Table~\ref{tab:sim_results}, we highlight the $e_{\text{peak}}$ that have an eccentricity difference smaller than $\pm0.02$ compared to our Model 2e. We find similar eccentricity values for all the disks with a planet, and also for the setups with no planet, with an aspect ratio of $h/r=0.05$ and binary eccentricity of 0.15 and 0.25. 
Previous studies had already determined that the disk eccentricities are the lowest in simulations where the binary eccentricity is $\approx0.16$ \citep[e.g.,][]{kley2014}; therefore, it is not surprising that some of the simulation setups with low $e_{\text{bin}}$ can reach eccentricities comparable to the setups with planet. 

% Disks with planets are more azimuthally symmetric
Even though our simulations show that low eccentricities can be achieved in circumbinary disks without the need of a planet, the azimuthal asymmetries in gas density are decreased by more than half when a single planet is introduced. This is another quantity that can be directly compared to our observations. The parametric Model 2e does not consider azimuthal variations in the intensity, therefore, we can use the amplitude between the most positive and most negative residuals as a reference approximation for the ratio between the brightest and dimmest parts of the ring peak. We calculate this value from the residuals imaged with a robust parameter of 0.0, shown in Fig.~\ref{fig:mod_two_ecc}. The residual image (after subtracting the best Model 2e) has the advantage that all the emission it contains corresponds to the non-axisymmetric emission from the disk and, therefore, it is better suited to quantify the azimuthal asymmetry. We find the ratio between the peak positive and peak negative residual to be $q_{\text{2e}}=1.09$. This a value is obtained from dividing two quantities in Jy\,beam$^{-1}$ units, consequentially, it should not be strongly dependent on the beam size or shape, although the brightness still has a preferred direction, parallel to the major axis of the beam. Any improvement to the parametric model would only result in a decrease in the amplitude of the residuals, which means that $q_{\text{2e}}$ is closer to an upper limit for the ring contrast. This ratio is not a density ratio as in the case of the simulations, since the brightness is also affected by optical depths effects even if we assumed a constant temperature. Nonetheless, this $q_{\text{2e}}=1.09$ is a reference for the extent to which the observation is axisymmetric, after correcting by eccentricity.

% dust overdensity should have a comparable amplitude to gas overdensity
A comparison between the density in the simulated gas and observed dust continuum requires an additional assumption, which is that the dust will have an enhancement in density of the same amplitude as the gas. The asymmetries observed in our simulated disks are not dust traps (e.g.,  vortex-like structures) and, rather, it is rather akin to a  "traffic jam"\ due to the eccentricity of the disk. In this scenario, \citet{ataiee2013} found that the azimuthal contrast in the dust can be as high as the contrast in the gas, since the precursor for the local density enhancement is the azimuthal difference in orbital velocities and not an azimuthal dust trap.

\begin{figure*}[t]
 \centering
        \includegraphics[width=15cm]{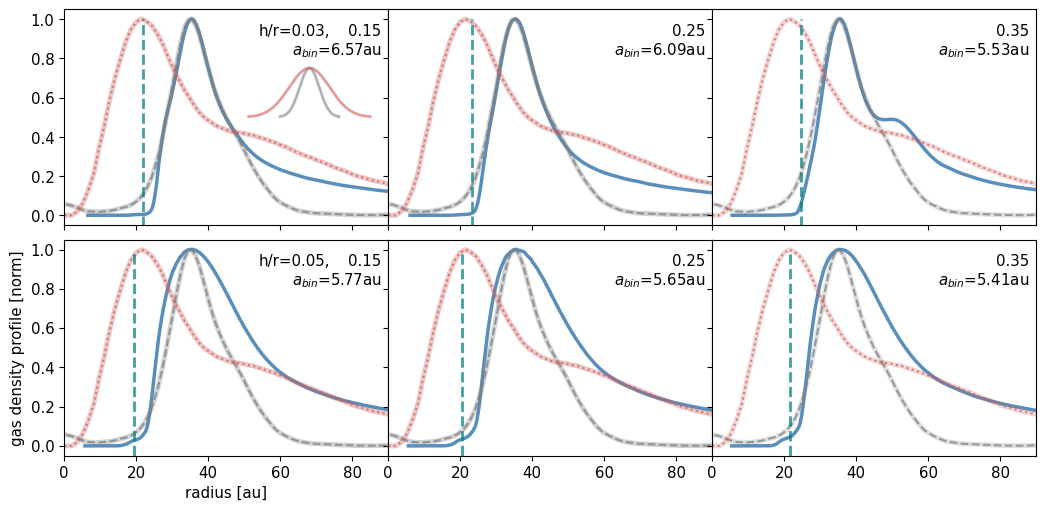}\\
        \vspace{-0.1cm}
   \caption{Comparison of azimuthally averaged radial profiles from the $^{12}$CO emission (in dotted red), dust continuum emission (in dashed gray), and gas density profile from each simulation setup (in solid blue), calculated as the median from the last 1000 orbits. The value of $a_{\text{bin}}$ is calculated for each simulation to match the peak density position with the peak brightness position of the dust continuum. The values for $e_{\text{bin}}$ increase from left to right, and each row has a constant disk aspect ratio. The average radial resolution of the $^{12}$CO and dust continuum are shown in the upper left panel, with Gaussians of the same colors. A dashed vertical line marks the position of the simulated planet.
   }
   \label{fig:prof_comp}
\end{figure*}

% Circumbinary disks azimuthal symmetry as a tracer of planet population
The azimuthal contrast along the peak density ring is about three to seven times smaller when a planet is included in the disk, as highlighted in Table~\ref{tab:sim_results}. The combination of low disk eccentricities and low azimuthal variations along the ring could be the key parameters to distinguish between disks that do or do not host a gap-opening planet inside the disk cavity. Observations with good S/N and high spatial resolution of circumbinary disks have been obtained for other systems, such as GG\,Tau\,A and AS\,205\,S, and both show azimuthal brightness variations of higher amplitude compared to the contrast detected in CS\,Cha. Increasing the sample of circumbinary disks with deep observations is needed to draw a definitive conclusion.

% The planet must allow material to get into the cavity
Another constraint that has to be taken into account to estimate the planet mass is the amount of material that it lets into the cavity through streamers. Our observations detect the presence of dust inside the cavity, which is likely to be part of the circumstellar disk of each star. Due to the faintness of this signal, we do not model it as two individual sources, but we rather use a single Gaussian to describe it. This emission is bright enough not to be neglected, as shown in the residuals from Models 2g and 3g in Fig.~\ref{fig:model_others}. In the context of circumbinary disks, these individual circumstellar disks seem to be brighter in younger systems such as GG\,Tau\,A or IRAS\,04158+2805 \citep[][ respectively]{phuong2020, ragusa2021}, and constraining their properties can give more insight into the processes shaping the cavity. High angular resolution observations in shorter wavelengths, such as ALMA Band 9 or Band 10, could have a better chance at detecting, at a higher S/N, the material around the stars and also the circumstellar material of CS\,Cha\,B, which is known to be brighter at shorter wavelengths \citep{haffert2020}.

\subsection{Cavity edge and ring morphology}

% Morphology of the disk changes for different setups
As shown in Figs.~\ref{fig:last_output} and~\ref{fig:prof_comp}, changing $e_{\text{bin}}$ and disk aspect ratios can induce different rings morphologies, as they differ in eccentricity, azimuthal symmetry, radial extent, and density distribution. Observationally constraining of the orbital parameters of the CS\,Cha binary stars will greatly reduce the degeneracy of the parameter space, as different binary eccentricities will not need to be sampled and the location of the ring peak and the ring width will also become quantities that can be compared among the observations and simulations. Additional observations of CS\,Cha in longer wavelengths such as $1.3$\,mm or $3\,$mm would allow us to test the azimuthal brightness variation along the ring in optically thinner emission, thus increasing the constrains in the proposed planet.

Observations of the disk in scattered light emission have found the cavity edge (if any) to be hidden by the coronagraph of SPHERE, therefore setting an upper limit of 15.6\,au \citep{ginski2018}. This value differs from the temperature peak at 21.6\,au that we observe in the $^{12}$CO emission, which is most likely the location of the gas cavity inner edge. The difference between those measurements is an additional constrain for the possible planet mass, and the disk physical conditions. For the same $e_{\text{bin}}$, different disk aspect ratios will also modify the shape of the ring inner edge, depleting the material closer or farther from the star, as shown in Fig.~\ref{fig:prof_comp}. To compare the observed profiles to the simulations, we scaled the value of $a_{\text{bin}}$ such that the peak density position matches with the peak brightness of the dust continuum. This scaling is made under the same assumptions discussed in Sect.~\ref{sec:disc_planet}, which is that the brightness peak of the dust continuum will trace the dust density maxima and it will coincide with the gas density maxima of the simulations.

The small $\mu$m-sized grains traced by the scattered light images could be getting through the planet orbit via streamers, which connect the main circumbinary ring to the binaries, and replenish their circumstellar disks. In fact, Fig.~\ref{fig:prof_comp} shows that the simulated gas is not always depleted at the planet location, specially in the setups with $h/r=0.05$. The small grains coupled to the gas at the planet orbit location could contribute to the difference in cavity size when observed with different tracers. Interestingly, when the semi-major axis of the planet is scaled from $a_{\text{bin}}$ to au, all of our simulations locate the planet almost at the same position as the $^{12}$CO peak brightness, which probably coincides with the cavity edge, where the $^{12}$CO reaches its highest temperature. Follow-up observations with alternative molecular lines, in combination with the accurate determination of the binaries orbits, would set even stronger constrains over the disk physical conditions and the candidate planet mass.

Finally, our work shows the feasibility of applying visibilities modeling with more than one eccentricity. The problem of analytically describing a coordinate system with variable eccentricity as a function of distance can be solved by approximating the emission with multiple components at different distances and eccentricities. While challenging, a combination of such approach with the visibilities modeling of the $^{12}$CO emission would overcome the limitations related to image reconstruction with synthesized beam convolution and possibly recover a precise description of the cavity inner edge morphology, location, and eccentricity, as we did for the dust continuum emission.

%%%%%%%%%%%%%%%%%%%%%%%%%%%%%%%%%%%%%%%%%%%%%%%%%%%%%%%%%%%%%%%%%%%%%%%

\section{Conclusions}
\label{sec:conclusions}

This work presents an analysis of the high angular resolution ($\approx 30\times46$\,mas) observations at 0.87\,mm of the CS\,Cha system, composed of a spectroscopic binary (usually referred just as CS\,Cha) and a co-moving companion at $1.3''$ known as CS\,Cha\,B \citep{ginski2018, haffert2020}. Our observations do not detect any significant emission from the expected position of CS\,Cha\,B, neither in the dust continuum emission or $^{12}$CO. We set an upper limit for its disk 0.87\,mm continuum emission to be $35.4\,\mu$Jy, which is the $3\sigma$ limit in the image generated with a robust parameter of 1.0.

The circumbinary disk resolves into a single ring, which has a peak in the dust continuum emission at $35\,$au and $22\,$au in the $^{12}$CO J:3-2 transition. Both the dust and gas emission show evidence of non-circular orbits, which is expected for circumbinary disks. The eccentricity in the dust continuum is constrained by visibility modeling, and we find the peak of the ring to have an eccentricity of $0.039$ and the contrast between the brightest and dimmest part along the peak ring to be at most $9\%$.

From our simulations of circumbinary disks, we find that including a Saturn-mass planet is in better agreement with the observations compared to the disk with no planet, as it can reproduce the low eccentricity and low azimuthal contrast over the ring. Even though it is possible to achieve low disk eccentricities without the need of a planet, the azimuthal symmetry of the disks is only achieved when a planet is present. Additional deep observations of other circumbinary disks could reveal if there is a difference within the circumbinary disks population between disks that do or do not host a gap-opening planet within the cavity.

The accurate determination of the orbital parameters of the CS\,Cha binary would unlock several additional observables that could be directly compared to the simulations, such as the ring location, ring morphology, and outer-disk radius, which are only measured in binary-separation units in the current simulations.

\section*{Acknowledgments}

We would like to dedicate this work to the memory of Willy Kley, who passed away during the development of this paper. We are grateful for his help and advice. 
The authors thank the anonymous referee for providing a constructive and detailed report.
N.K. and P.P. acknowledges support provided by the Alexander von Humboldt Foundation in the framework of the Sofja Kovalevskaja Award endowed by the Federal Ministry of Education and Research.
Anna Penzlin was funded by grant KL 650/26-2 from the German Research Foundation (DFG). This project has received funding from the European Research Council (ERC) under the European Union’s Horizon 2020 research and innovation programme (grant agreement No. 101002188). 
L.P. and A.B acknowledges support by ANID, – Millennium Science Initiative Program – NCN19\_171. A.B. also acknowledges ANID BASAL project FB210003 and Fondecyt (grant 1190748), and L.P. gratefully acknowledges support by the ANID BASAL projects ACE210002 and FB210003.
This paper makes use of the following ALMA data: ADS/JAO.ALMA\#2017.1.00969.S. ALMA is a partnership of ESO (representing its member states), NSF (USA) and NINS (Japan), together with NRC (Canada), MOST and ASIAA (Taiwan), and KASI (Republic of Korea), in cooperation with the Republic of Chile. The Joint ALMA Observatory is operated by ESO, AUI/NRAO and NAOJ.

\bibliographystyle{aa}

\begin{appendix} 
\section{Additional figures \label{app:figures}}

\begin{figure}
 \centering
        \includegraphics[width=18cm]{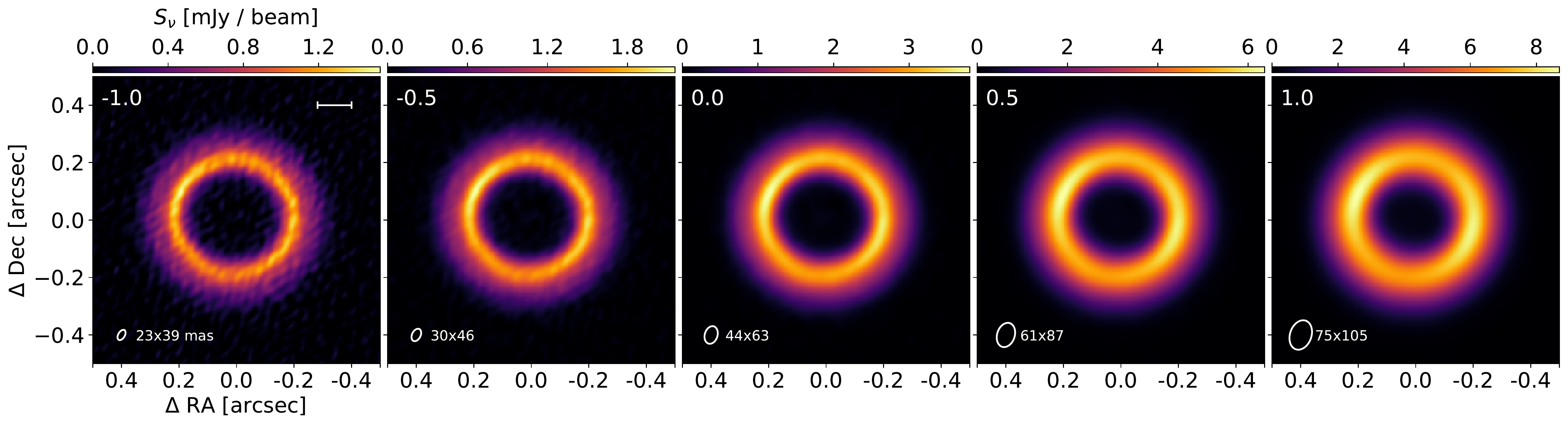}\\ \vspace{-0.2cm}
   \caption{CS\,Cha dust continuum emission, as imaged with different robust parameters. The size of the synthesized beams are shown in the bottom left corner of each panel.}
   \label{fig:continuum_gallery}
\end{figure}

\begin{figure*}
 \centering
        \includegraphics[width=18cm]{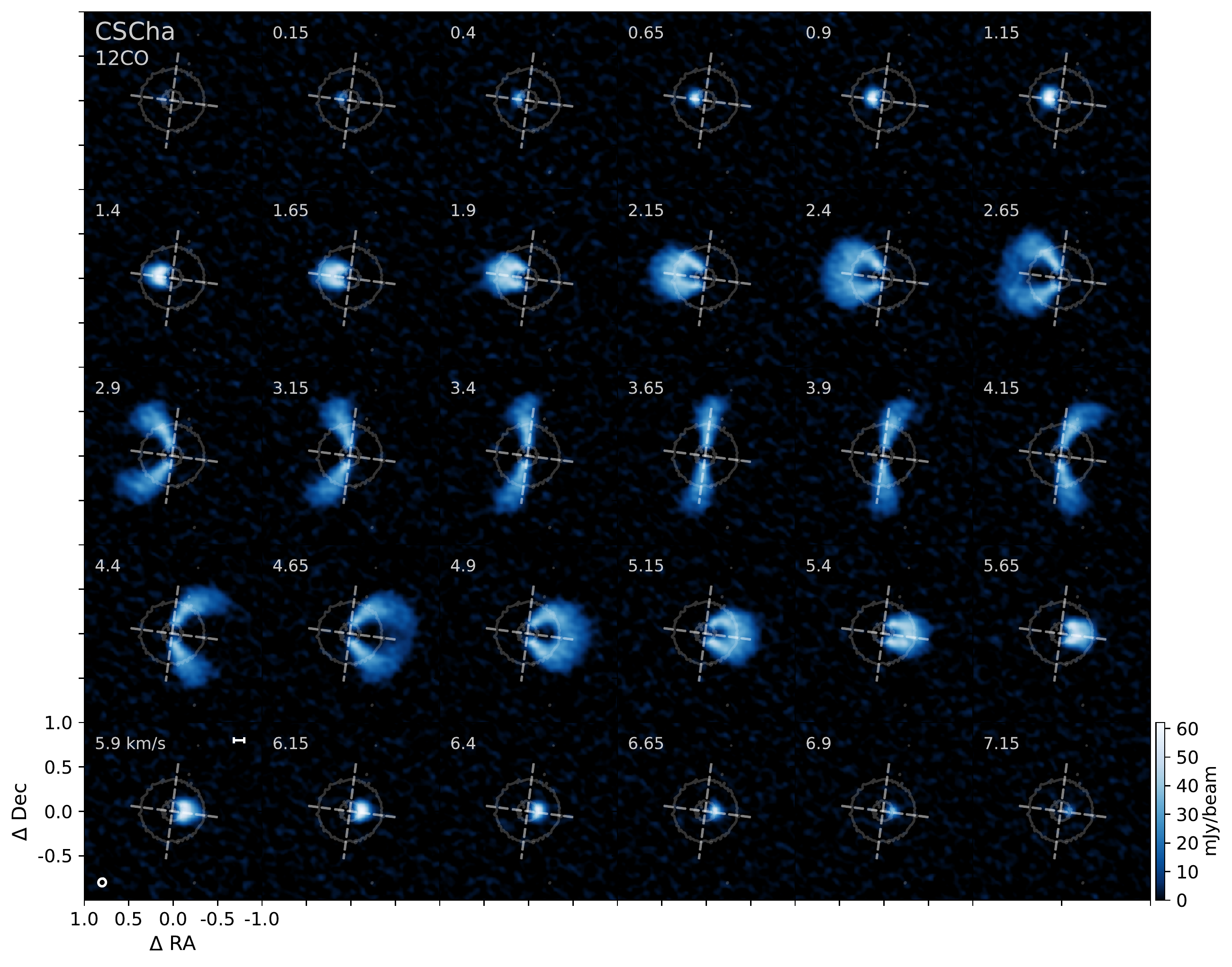}\\ \vspace{-0.1cm}
   \caption{$^{12}$CO Channel maps of CS\,Cha, generated with a robust parameter of 0.0. The velocity of each channel is shown in the upper right corner. The contours are the $5\sigma$ level of the continuum image generated with a robust parameter of 0.0. Lower left panel: Scale bar represents 20\,au at the distance of the source, and ellipse represents the beam size for all the images.}
   \label{fig:channel_maps}
\end{figure*}

\begin{figure}
 \centering
        \includegraphics[width=8cm]{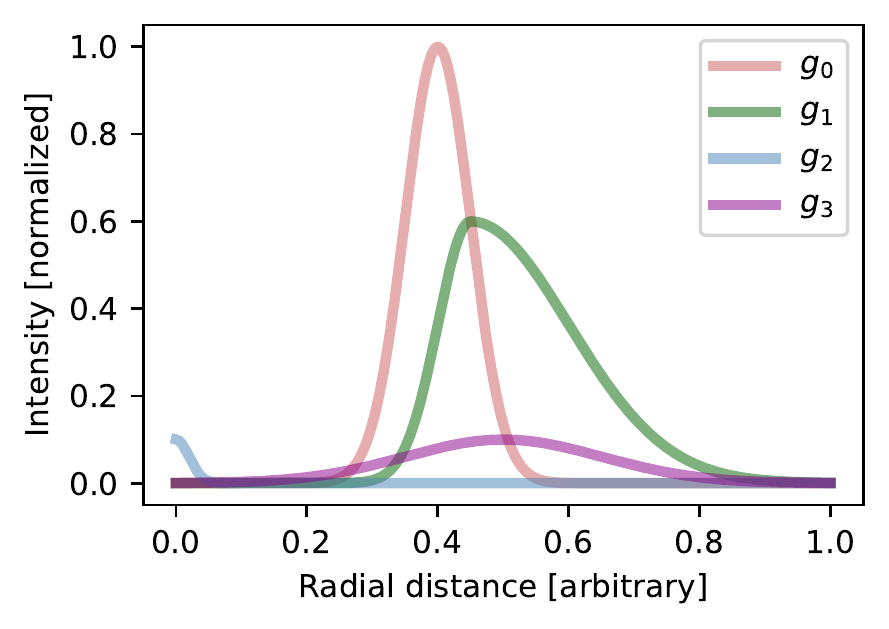}\\ \vspace{-0.2cm}
   \caption{Schematic profile of the components considered in our uv modeling.}
   \label{fig:schematic_profiles}
\end{figure}

\begin{figure*}
 \centering
        \includegraphics[width=16cm]{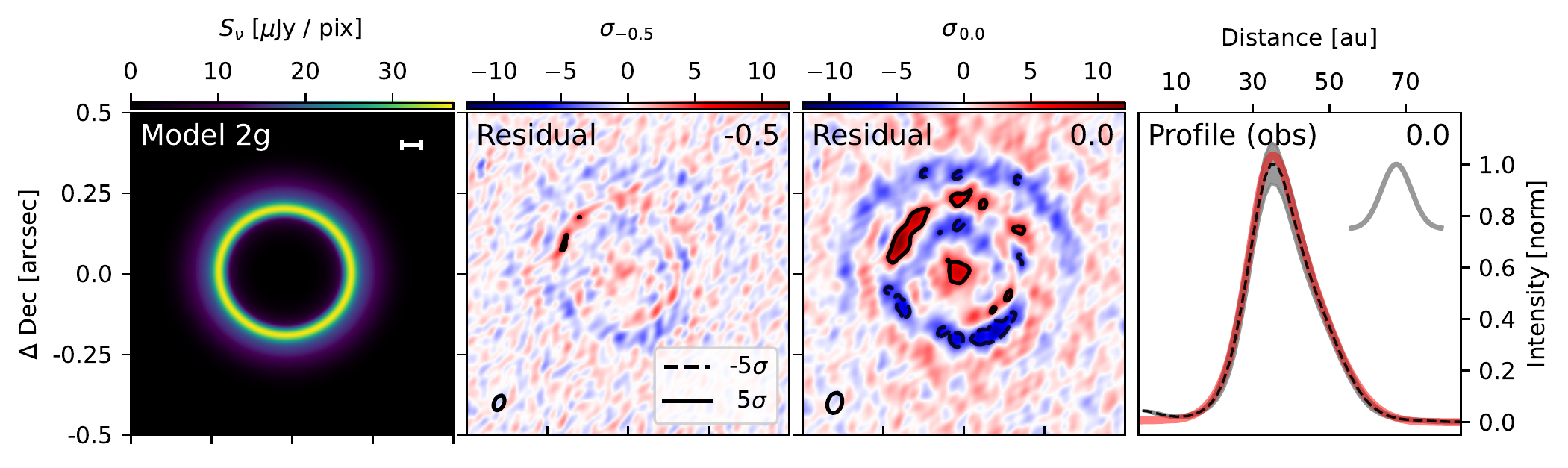}\\ \vspace{-0.1cm}
        \includegraphics[width=16cm]{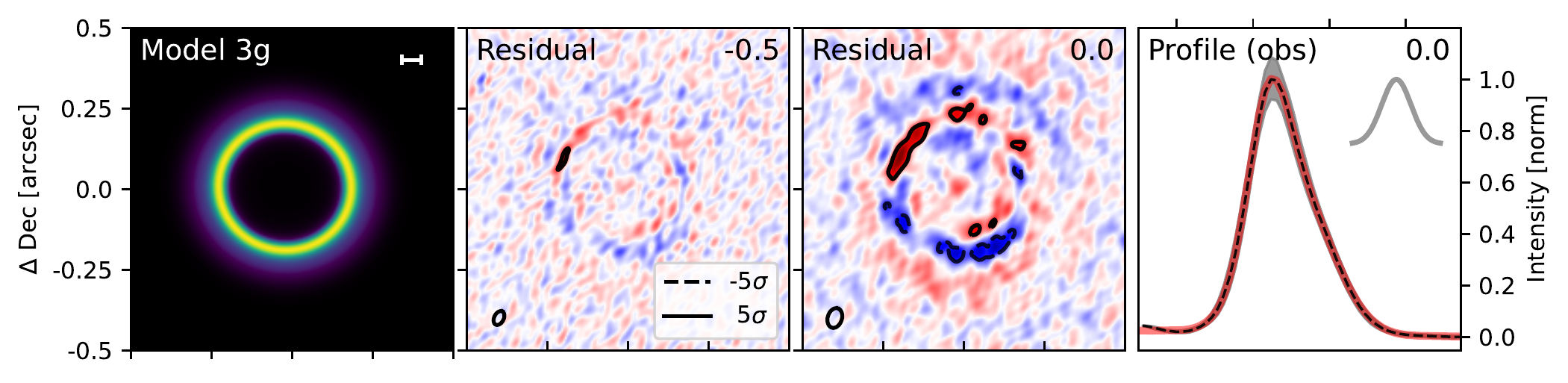}\\ \vspace{-0.1cm}
        \includegraphics[width=16cm]{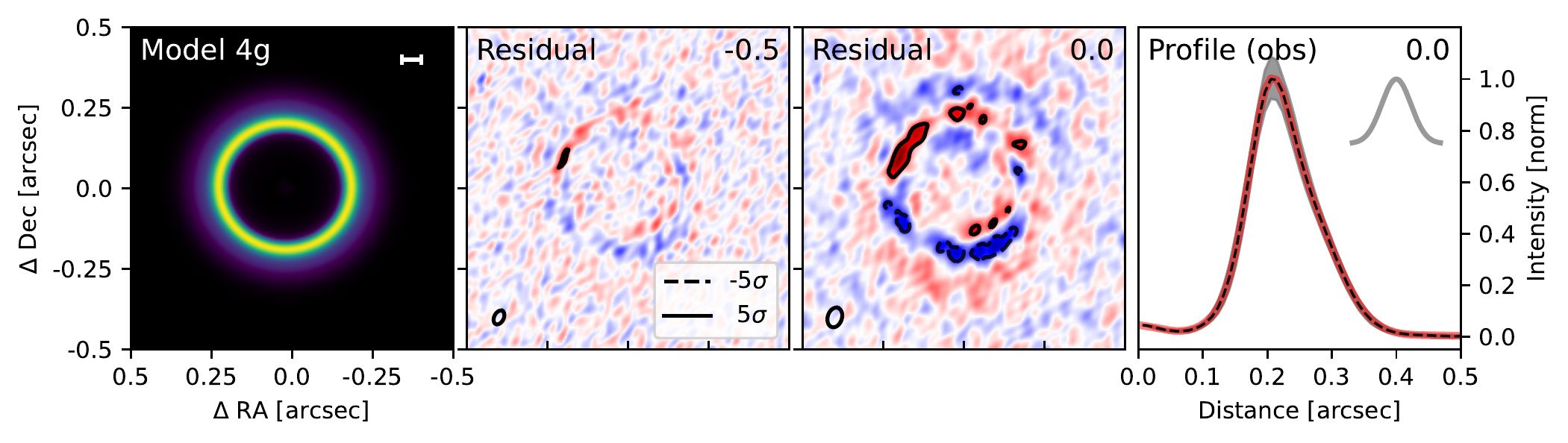}\\ \vspace{-0.1cm}
   \caption{Best solutions for the dust continuum emission generated with the Models 2g, 3g and 4g. Left panel shows the best model, and middle panels shows the residuals left by the best model after being imaged with different robust parameters, shown in the upper right corned. Right panel shows the intensity profile of the model obtained from \texttt{tclean} (in dashed black) and the best respective model (in red).}
   \label{fig:model_others}
\end{figure*}

\end{appendix}

\end{document}